\documentclass[11pt]{article}
\usepackage[utf8]{inputenc}
\usepackage{hyperref}
\usepackage{pslatex}
\usepackage{graphicx}
\usepackage[T1]{fontenc}
\usepackage{subfig}
\usepackage{threeparttable}
\usepackage{booktabs}
\usepackage{floatpag}
\usepackage{blindtext}
\usepackage{amsmath}
\usepackage{etoolbox}
\usepackage{pdflscape}
\usepackage{array}
\usepackage{float}
\usepackage{soul}
\usepackage{color}
\usepackage{amssymb}
\usepackage{rotating}
\usepackage{caption}
\usepackage{needspace}
\usepackage{multirow}
\usepackage{mathabx}
\usepackage{amsthm}
\usepackage{placeins}
\usepackage{setspace}
\usepackage{authblk}
\usepackage{comment}
\usepackage{epstopdf}
\usepackage{layout}
\usepackage{layouts}
\usepackage{authblk}
\usepackage[margin=0.92in]{geometry}
\newcolumntype{L}[1]{>
{\raggedright\let\newline\\\arraybackslash\hspace{2pt}}m{#1}}
\newcolumntype{C}[1]{>{\centering\let\newline\\\arraybackslash\hspace{2pt}}m{#1}}
\newcolumntype{R}[1]{>{\raggedleft\let\newline\\\arraybackslash\hspace{2pt}}m{#1}}

\usepackage{xr}

\usepackage[margin=0.92in]{geometry}

\makeatletter
\newcommand*{\addFileDependency}[1]{
  \typeout{(#1)}
  \@addtofilelist{#1}
  \IfFileExists{#1}{}{\typeout{No file #1.}}
}
\makeatother

\title{A recommender-network perspective on the informational value of critics and crowds}

\author[1]{Pantelis P. Analytis}
\author[2]{Karthikeya Kaushik}
\author[3]{Stefan M. Herzog} 
\author[4]{Bahador Bahrami}
\author[5]{Ophelia~Deroy}
\affil[1]{Department of Business and Management, University of Southern Denmark}
\affil[2]{Department of Psychology, UC Berkeley}
\affil[3]{Adaptive Rationality Center, Max Planck Institute for Human Development}
\affil[4]{Department of General Psychology and Education, Ludwig Maximilian University of Munich}

\affil[5]{Department of Philosophy, Ludwig Maximilian University of Munich}

\date{\today}
\begin{document}

\maketitle


\begin{abstract}

How do the ratings of critics and amateurs compare and how can they be combined? Previous research has produced mixed results on the first question, while the second remains unanswered. We have created a new dataset with wine ratings from critics and amateurs, and simulated a recommender system using the weighted $k$-nearest-neighbor algorithm. We then formalized the advice-seeking network spanned by that algorithm (i.e., who advises whom?) and studied people’s relative influence. We find that critics are more consistent than amateurs, and thus their advice is more predictive of people's taste than advice from amateurs. Getting advice from both groups can further boost performance, but only by a small margin. Our network-theoretic approach allows us to identify influential critics, talented amateurs, and the information flow between groups. Our results provide evidence about the informational function of critics, while our framework is broadly applicable and can be leveraged to devise good decision strategies and more transparent recommender systems.

\end{abstract}
\vspace{3.5mm}

\enlargethispage{3mm}

\newpage

%

\newpage 
\vspace{12mm}

\section{Introduction}
Whether it is a film currently in cinemas,  a restaurant that just opened, or vintage wine, people like to voice their judgments on matters of taste. Even more so, for a few select individuals---critics---expressing judgments on matters of taste has turned from a fun past time to a profession. Critics are employed in the daily and weekly press to assess restaurants, theater, movies, and wine labels, while some even run popular television shows. Critics may function as information producers, supplying readily accessible information about products to the wider public \cite{hsu2012evaluative,sharkey2022expert}. This key function of critics is especially pronounced on websites such as Metacritic and Rotten Tomatoes, which have based their business model on collecting critics' opinions and devising scores that summarize their opinions, or when they are hired as judges in prestigious competitions. So far, it has been shown that the judgments of influential critics can predict \cite{eliashberg1997film} or even alter the overall financial performance of products \cite{ali2008impact,basuroy2003critical}. But are the evaluations of critics more informative than those of amateur consumers, and if so, why?

Even though there is clear evidence for the sizable market impact of at least some critics, the informational value of their opinions for the broader public has often been contested. Popular lore and scholars in different traditions have argued that in matters of taste, there is often a big, unresolvable divide between critics and the general public \cite{gans2008popular}. Some would suggest that critics and other experts engage in hairsplitting over nuances that are often lost on the rest of the world, who consider expert opinions high-brow, and ivory tower, if not altogether imaginary or downright boring \cite{hekkert1996beauty,bourdieu2012distinction}. Others emphasize that general audiences lack the ability to assess sophisticated (cultural) products, and only people who have trained their senses are capable of doing so \cite{adorno1997aesthetic}.  Regardless of the point of view adopted, other people may rarely care about what critics think, whatever theirs standards of taste may be. Thus, although critics's opinions may be impactful (e.g. because they bring attention to certain categories of cultural or experience products \cite{shrum1991critics}), they could also be less informative than crowd-sourced aggregate scores, the opinions of representative individuals in the crowd or even a randomly selected person from the street \cite{holbrook1999popular}. That is, the opinions of a critic or a group of critics might be less correlated with and, most importantly, less predictive of the tastes of people in the wider  public, than those of other amateurs.

%
A number of past studies in marketing, psychology, sociology, and cultural economics have explored whether the opinions of critics and amateur consumers correlate in different taste domains and have produced mixed results \cite{holbrook1999popular,holbrook2005role,wanderer1970defense,hirschman1985relationships,lundy2019good,boor1992relationships}. Holbrook, for example, finds modest correlations between the tastes of critics and the general public in the case of movies \cite{holbrook1999popular,holbrook2005role}. Other studies looking at Broadway productions, popular music, and plays have found slightly negative  \cite{hirschman1985relationships,lundy2019good} or even strong positive \cite{wanderer1970defense} correlations between critics and the wider audience. Taken at face value, the previous literature remains inconclusive, and even if most studies seem to suggest that critics opinions relate to some extent to those of the wider public, it is still unclear whether critics' opinions are more informative than those of amateurs. 
Furthermore, although it is obvious that there is substantial variation in the extent to which different critics or amateurs can influence or inform others \cite{cameron1995role}, most existing studies comparing the tastes of critics and amateurs disregard it and focus on reporting average correlations between the two groups. As a consequence, there are no methods for identifying the best critics to follow in a certain domain. Last, rather than focusing on drawing advice from specific individuals or groups, it would be valuable to know when it is beneficial combine advice from different critics and/or amateurs. 

Several questions about the ways in which critics and/or amateur crowds can inform and influence the broader public remain unanswered. First, are the opinions of critics more informative than those of amateurs/amateur crowds, and if so, why? Second, is it possible to identify the most informative critics and talented amateurs who have the potential to become critics? Third, how can the opinions of critics and amateur crowds be best combined and how would information flow between the two groups in an efficient advice network? To answer these questions we need a methodology that would allow us to measure the informational value of the opinions of different individuals or groups for others as well as the their potential to influence them.

We put forward an approach that combines methods from the recommender systems \cite{breese1998empirical,amatriain2009wisdom}, machine learning \cite{csimcsek2013linear} and network science communities \cite{krackhardt1987cognitive,currarini2009economic} and from the study of expert judgments in matters of fact, where different strategies for choosing among experts or identifying the best experts have been explored in domains ranging from medicine to agriculture \cite{shanteau1992competence,ashton1986combining,kurvers2019detect}. We leverage the \emph{weighted k-nearest neighbors} algorithm (k-nn), a classic recommender systems algorithm that encodes an array of strategies for choosing among experts (and amateurs) as special cases  \cite{analytis2018social}
(see also \ref{tab:strategyTable}).  For each individual and item, our implementation of the algorithm draws advice from the $k$ most similar other individuals in the database who have evaluated that item and weights their opinions according to a similarity sensitivity parameter to form a prediction of how much the target individual will like the item. Thus, the main difference between \emph{k-nn} and previously proposed strategies for choosing among experts is that weighted \emph{k-nn} relies on observed similarity between the target individual and potential advisers in past ratings rather than the observed performance in predicting some quantity of interest. By assessing the out-of-sample performance of the algorithm when drawing advice from critics, amateurs, or both groups and varying the number of neighbors $k$ and the similarity sensitivity parameter $\rho$, we can compare the predictive performance of different strategies involving each or both of these groups for each person, and thus assess their relative informational value at the individual and aggregate level.

The \emph{k-nn} algorithm spans an advice network among different individuals (i.e., who advises whom?) \cite{analytis2020structure}, and it is thus possible to visualize and study the properties of such a ``taste network'' in any dataset where a group of people have evaluated a set of items, even when the overlap in the ratings of different people is relatively small (i.e., sparse rater--item matrices). Individuals (advisors) whose tastes appear relevant for many similar others (advisees), be they critics or amateurs, are sought often for their advice from the  \emph{k-nn} algorithm---they have a large \emph{recommender potential}. This potential, however, can only materialize as \emph{recommender influence} when the advisers have experienced the items that their advisees are considering, and can thus supply their ratings when the algorithm (or an advisee) seeks it. To assess the flow of influence between two categories of individuals---critics and amateurs---we adapt the notion of homophily from network science and show how it can be applied to domains of taste. The concept of \emph{taste homophily} is general and can be applied to any rating dataset where items have been evaluated by two or more categorical groups of raters. Thus, our methods be applied by social and behavioral scientists in other taste domains, and they can also be used by the recommender systems community to improve the transparency and interpretability of the recommendation process.

Since the advent of the internet and social media, the informational landscape in the wine world has been changing swiftly. Vivino has emerged as a reliable alternative to critics for getting access to information about the quality of different wines, and as with other opinion aggregation websites (e.g. IMDB, Yelp, Amazon) it gives wine consumers a way to consult a single information source to inform their choices,  bringing forth a democracy of taste where each individual in the user base can contribute to the aggregate wine ratings \cite{kopsacheilis2023crowdsourcing}. Personalization algorithms can search for similarity patterns between one person and others in rating databases, and they can receive recommendations based on a number of apparently similar people from across the globe, thus circumventing critics altogether \cite{adomavicius2005toward,resnick1997recommender}. To examine how the opinions of critics relate to those of amateurs we created a new dataset consisting of the ratings from both renowned wine critics and regular wine consumers (i.e., amateurs, non-professionals). We obtained critics' data from Bordoverview, a website summarizing the en primeur ratings (first sampling of a production year) on Bordeaux wines, and matched these ratings with amateur data on the same wines from Vivino. The resulting dataset has the properties needed to assess how wine critics and amateurs can inform and influence others and how their judgements on matters of taste can be best combined.

\section{Methods}

\textbf{Bordeaux wine dataset} We first obtained ratings by professional critics from Bordoverview and ratings by amateurs from Vivino by scraping the two websites. We then combined the two datasets and restricted our analyses to wine labels that were included in the Bordoverview list and had at least 5 reviews in Vivino and to Vivino users with more than 50 ratings in the resulting wine label list. This resulted in a dataset comprised of 1978 wine labels (322 different wines across 15 different vintages from 2004 to 2018), 14 professional critics (or wine magazines or other outlets employing critics), and 120 Vivino amateurs. The dataset has 25,907 ratings in total, and average density (i.e., mean proportion of rated wine labels relative to all wine labels) in the dataset is 4.7\% for amateurs and 53.3\% for professional critics. Ratings in both datasets were normalised within each rater using $z$-scoring, that is, transformed ratings now indicate for each rater how many standard deviations a rating was above or below the mean rating of that rater. This was done to make the ratings more comparable across the different rating scales that different critics use (e.g., 10 to 20 for Jancis Robinson vs. 75 to 100 for Jeff Leve) and the 1-to-5 scale used by Vivino. 

\textbf{Recommendation algorithms} In our analysis, we rely on the well-established $k$-nearest neighbors algorithm (\emph{k-nn}) \cite{resnick1994grouplens,resnick1997recommender,ekstrand2011collaborative}, that seeks the $k$ most similar individuals, allowing for differential weights \cite{breese1998empirical,nosofsky1992similarity}. 

Such a weighted nearest neighbor algorithm can be expressed as follows: 
\begin{equation}
\widehat{u_m} =  \frac{1}{\sum_{j=1}^kw_j}\sum_{j=1}^kw_j\times u_j 
\label{eq:knn}
\end{equation}
where $\widehat{u_m}$ is the estimate of the utility of an item $m$ for the target individual, $j$ is the $j$th nearest neighbor to that target individual, and $w_j$ the weight put on that other individual. For $k=1$, the algorithm seeks advice from only the most similar other individual. Setting $k = N-1$, where $N$ is the total number of individuals in a dataset, amounts to the weighted averaging strategy. For values of $k$ between these two extremes, we obtain the usual \emph{k-nn} implementation,  with differential weights.

We used the Pearson correlation coefficient as a measure of similarity ($w$) between two individuals $i$ and $j$ \cite{herlocker2004evaluating}, defined as follows: 
\begin{equation}
  w(i,j) =  \frac{\sum_{m=1}^{M}(u_{im} - \widebar{u_{i}})(u_{jm} - \widebar{u_{j}})}{\sum_{m=1}^{M}\sqrt{(u_{im} - \widebar{u_{i}})^2(u_{jm} - \widebar{u_{j}})^2}}
  \label{eq:correlation}
\end{equation}

where $u_{im}$ is the rating that the target individual $i$ gave to item $m$ and $u_{jm}$ is the evaluation that the $j$th individual gave to the same item $m$. $M$ stands for the total number of items.

We use a similarity sensitivity parameter $\rho$ that allows us to amplify or dampen the weights $w(i,j)$ of other people \cite{breese1998empirical}. We directly modify the weights obtained from Eq. \ref{eq:correlation} using the following scheme: 

\begin{equation}
w'_j = \begin{cases}
w_j^\rho \: \: if \: \: w(i,j) \geq 0 \\
0 \: \: otherwise. 
\end{cases}
\label{eq:amplification}
\end{equation}
\\

By varying $k$ and $\rho$, we can produce several social learning and information aggregation strategies studied in the behavioral and management sciences (Table \ref{tab:strategyTable}; also see \cite{analytis2018social,analytis2020structure}), some of which have been used to study how to best combine expert opinions (i.e. forecasters, doctors, etc., see \cite{ashton1986combining}). For instance, setting $\rho = 0$ and $k = N-1$ produces the unconditional wisdom of the crowds strategy \cite{einhorn1975unit}. Setting $\rho = 0$ and $k = n$ gives the original unweighted formulation of \emph{k-nn}, which corresponds to the select crowd strategy in psychology and management \cite{mannes2014wisdom}. Setting $\rho > 0$ weights the opinions of people more similar to the target individuals. This is common in implementations of the nearest neighbors strategy in collaborative filtering \cite{breese1998empirical} and strategies aggregating opinions of more competent experts in management \cite{budescu2014identifying}. As $\rho$ increases, the weight distribution becomes more unequal and the most similar individuals have a proportionally larger weight. In addition to applying \emph{k-nn} to our entire dataset we applied the algorithm to subsets of the data, giving it either access to only the ratings of critics or amateurs (i.e. searching for the $k$ most similar critics or amateurs). 


We took two measures to make our analysis routines robust  to inconsistencies that could be produced due to the sparsity of the dataset. First, we only considered correlations when the number of overlapping items between target individual $i$ and adviser $j$ is higher than 5 and set all the remaining correlations to the mean correlation between individual $i$ all other individuals $j$ with whom they had an overlap of more than 5 items. Using such thresholds (or other methods of discounting observed correlations from sparse data) is a common approach when deploying collaborative filtering algorithms. Further, when some of the $k$ most correlated individuals had not evaluated the target item, the algorithm searched further down the list of others ranked by similarity until a committee of $k$ people was formed or there were no further potential advisors (in which case the committee had fewer than k advisers). This is a less common implementation of the \emph{k-nn} algorithm, but suitable for the size of our dataset and our research objectives.

\textbf{Performance of \emph{k-nn}} We assessed the out-of-sample performance of the weighted \emph{k-nn} algorithm by consistently leaving out 10 items for each individual and using the remaining items to learn the correlations between individuals. This approach is a variation of the leave-one-out approach in recommender systems \cite{christakopoulou2016local}, and ensures that that there are sufficient pair comparisons to be predicted in the test set per individual. The correlations learned in the training set were then used to find the $k$ most similar individuals to the target individual who have evaluated the item and were then up(down)-weighted according to $\rho$ (see Eqs. \ref{eq:knn} \& \ref{eq:amplification}). Thus, for each target individual and item, $k$ other individuals (whose opinions where weighted according to $\rho$) were used to predict how much the target individual would like that item. We repeated this process 1000 times and averaged the results across repetitions. As a measure of performance we used the number of correct decisions made by the model when choosing between a pair of items in the test set (45 choices in total, ties were resolved at random). We opted for this measure because it is intuitive and easily communicable  (1 corresponds to perfect choices and 0.5 correspond to random choices, also see \cite{analytis2018social,analytis2023collaborative}). It has been used before to assess the performance of decision or inference strategies in psychology and the management sciences \cite{gigerenzer1996reasoning} and it closely corresponds to the number of concordant pairs measure used by the recommender-systems community \cite{koren2011ordrec}.

\begin{table*}[ht!]
\vspace{1mm}
\setlength\tabcolsep{1.5pt}
\begin{tabular}{llll}
\hline
 \textbf{Social learning (taste)} & \textbf{Social learning (objective)}    &  \textbf{Algorithm parameters}                                                                           &  \textbf{Cognitive strategies} \\ \hline
 Doppelg\"{a}nger    \cite{Yaniv2011receiving,analytis2018social}        &  Follow the expert \cite{laland2004social} & $k = 1$ and $\rho $ = any & Take the best \cite{gigerenzer1999simple,hogarth2005ignoring}    \\ 
 Clique   \cite{muller2017wisdom,Yaniv2011receiving}     &  Select crowd  \cite{mannes2014wisdom,goldstein2014wisdom}   &  $k = n$ and $\rho = 0  $ &  -- \\ 
 Weighted clique         &  Weighted select crowd  \cite{tetlock2016superforecasting}    & $k = n$ and $\rho >  0  $                                &  --    \\ 
 Weighted crowd  \cite{analytis2018social}         &  Weighted crowd   \cite{budescu2014identifying}      &  $k = N - 1$ and $\rho > 0$                               &  Weighted additive  \cite{payne1993adaptive,dawes1974linear}   \\ 
 Whole crowd  \cite{Yaniv2011receiving,analytis2018social}         &   Averaging  \cite{hogarth1978note}      &  $k = N - 1$ and $\rho = 0$                               & Equal weights  \cite{dawes1974linear,einhorn1975unit}  \\ \hline
\end{tabular}
\vspace{1mm}
\caption{Correspondence between the collaborative filtering algorithm parameterizations we consider (see Equations \ref{eq:knn}, \ref{eq:correlation}, and \ref{eq:amplification}) and the social learning and information aggregation strategies broadly studied in the social and behavioral sciences \cite{analytis2020structure}}  
\vspace{-2mm}
\label{tab:strategyTable}
\end{table*}

\enlargethispage{3mm}

\textbf{Reconstructing the \emph{k-nn} advice network}\label{sec:networks} We studied the advice network spanned by \emph{k-nn} \cite{lathia2008knn,analytis2020structure} by constructing advice networks---for different values of $k$ and $\rho$---with nodes representing the different individuals in the dataset and directed edges representing an individual seeking advice from other individuals (or more precisely, \emph{k-nn} seeking advice on their behalf). While all individuals had by definition the same number of $k$ outgoing edges connecting them to other nodes, people could have a varying number of incoming edges depending on how often \emph{k-nn} sought their advice for other individuals. We used node strength, defined as the sum of weights of the incoming edges as  as a measure of social influence that naturally fits the weighted \emph{k-nn} algorithm and weighted networks more generally \cite{barrat2004architecture}. For $\rho = 0$ this measure collapses to in-degree. We then measured the \emph{recommender potential} of each individual, by calculating the node strength resulting from the \emph{k-nn} algorithm when disregarding missing values. That is, we counted only how often people were in the first $k$ individuals sought by the algorithm and their relative weights in the committees formed (expressed as a proportion), regardless of whether they had a rating to contribute for the item in question. We then calculated the \emph{recommender influence} of an individual by computing who actually contributed to recommendations and how much so. That is, we calculated how often our implementation of the \emph{k-nn} algorithm sought and used advice from an individual to predict how much another person would like an item, and the relative weight of such advice in the committees that eventually formed. Note that these two metrics of influence converge to the same metric for full (i.e., non-sparse) datasets. Following previous analyses we report results averaged across 1000 repetitions of the simulation.

\textbf{Calculating taste homophily}: We calculate taste homophily in the two groups by adapting the homophily index used in the work of Currarini \cite{currarini2009economic} to weighted networks and then using a baseline that is appropriate for sparse data structures. We define $N_{i}$ as the number of individuals of type $i$ in the population and $N$ the total population. Similarly, we define $R_{i}$ as the number of ratings  contributed from individuals of type $i$ and $R$ the total number  of ratings. Then $p_{i} = N_{i}/N$ is the proportion of individuals of type $i$ in the population  and  $r_{i} = R_{i}/R$ is the proportion of their ratings. We will use these two measures as baselines to mark whether the tastes of a group are characterized by homophily. The homophily index $H_{i}$ then is defined as the proportion of weights $s_{i}$ directed to members of the same group, divided by the total sum of weighted nodes (both weights directed to the same and to different groups, $d_{i}$), that is,  $H_{i} = \frac{s_{i}}{s_{i} + d_{i}}$. A group is characterised by taste homophily in respect to their proportion in the population if $H_{i} > p_{i}$ and in respect to the ratings they have contributed if $H_{i} > r_{i}$. The first baseline is similar to standard definitions of homophily in the network science literature, but the second one also takes into account data density disparities in the considered groups --- a feature of most real world recommender systems. The same definitions can also be used at the individual level, to evaluate whether specific individuals draw information from people belonging in the same group or from people outside their group. Note that the homophily index can be also calculated taking into account only the initial calls of the algorithm, as with recommender potential. In that case only the relation $H_{i} > p_{i}$ is relevant (we perform this analysis in the Supplementary Material).

\section{Results}

\begin{figure*}[tbh!]
\begin{center}
\includegraphics[width=1\textwidth]{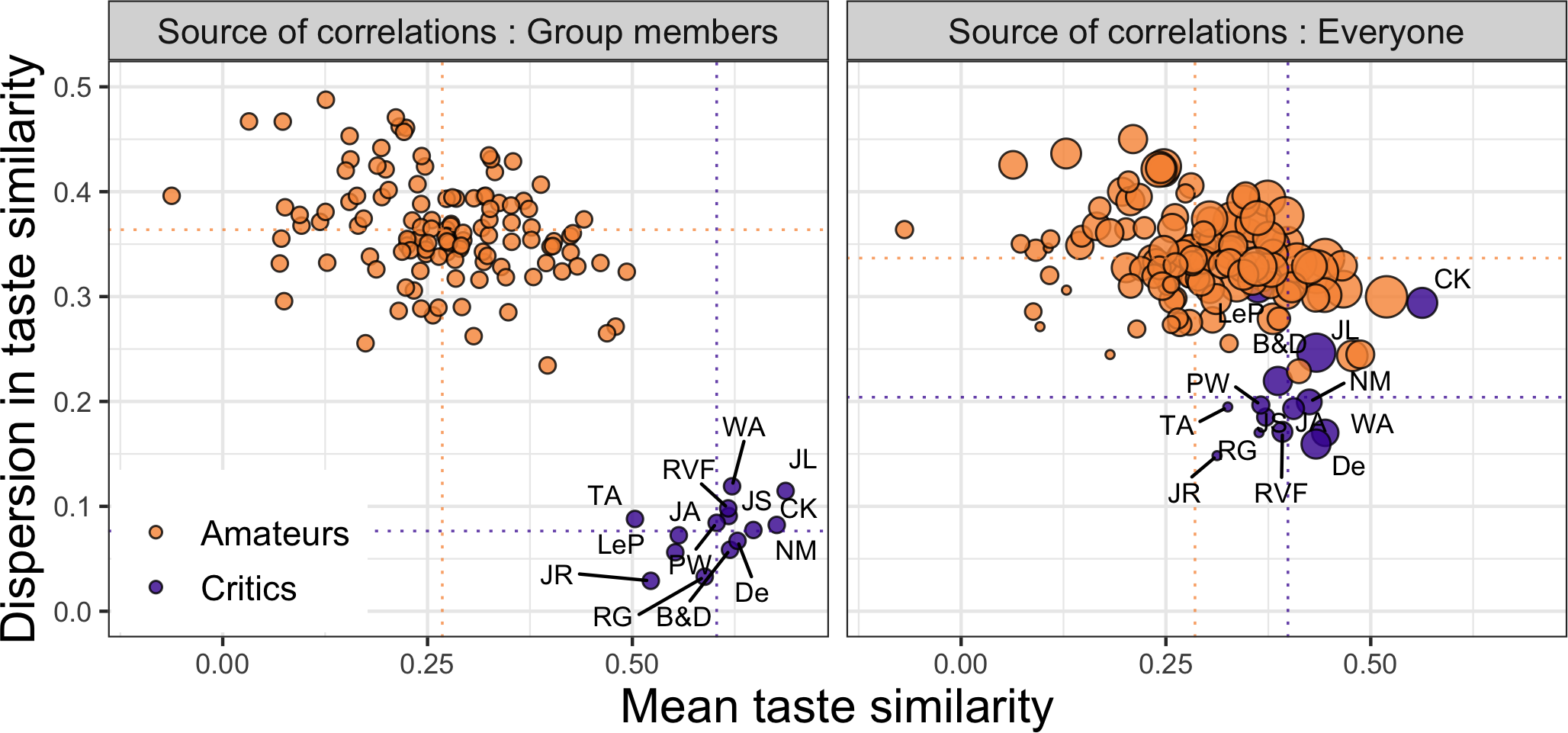}
\caption{\textbf{Intercorrelations with members of the same group and all other individuals.}  \textbf{Left}: The position of 14 professional critics and 120 amateurs on a 2-dimensional plane defined by mean taste similarity (i.e., mean correlation) and dispersion in taste similarity (i.e., SD of correlations) with members of the same group. \textbf{Right}: The position of the same 14 professional critics and 120 amateurs on the same plane, but this time with taste similarity calculated across all individuals (professional critics and amateurs). The color in both panels indicates whether an individual is a professional critic or an amateur and the point size in the right panel indicates recommender potential.  The dotted orange and purple lines indicate the average correlations and dispersion in taste similarity for amateurs and critics.  Only correlations when two individuals had an overlap of more than 5 ratings were considered in this graph. \textbf{Initials of professional critics}: WA --- Lisa Perotti Brown, NM --- Neal Martin, JR --- Jancis Robinson, TA --- Tim Atkin, B \& D --- Michel Bettanne and Thierry Desseauve, JS --- James Suckling, JL --- Jeff Leve, De --- Steven Spurrier, James Lawther, Beverley Blanning and Jane Anson, RVF --- Olivier Poels, Hélène Durange, and Philippe Maurange,  JA --- Jane Anson, LeP --- Jacques Dupont, PW --- Ronald DeGroot, RG --- Rene Gabriel, and CK --- Chris Kissack.
}

\label{fig:winePlane}
\end{center}
\end{figure*}

\textbf{Critics are more consistent than amateurs} In line with previous findings on the judgement consistency of experts vs. non-experts in matters of fact \cite{ashton2012reliability,shanteau1992competence}, we find more agreement among professional critics than among Vivino amateurs. The average taste similarity (correlation) among critics is 0.60, whereas the average taste similarity  among amateurs is 0.27 (see Figure \ref{fig:winePlane} left panel). A similar result also emerges when combining the critics and amateurs datasets into a single dataset and then calculating correlations across all individuals (i.e., irrespective of group membership; see Figure \ref{fig:winePlane} right panel). The average correlation between critics and everybody else is 0.39, whereas the average correlation between amateurs and everybody else is 0.29. These results are preserved even if we randomly remove ratings from the critic group to equate the average density of the two groups (see Figure \ref{fig:correlationDensitySparse} in the Supplementary material).

\textbf{Critics tend to be representative of the amateur audience}  Critics also tend to be more representative of the amateur population than other amateurs. The average similarity between critics and amateurs is 0.36, which is substantially higher than the average similarity among amateurs, which is 0.27. Nonetheless, for most amateurs the highest encountered correlations in the dataset are with other amateurs. There are two reasons for that: first, there is a sizable sub-group of fifteen to twenty amateurs whose average correlations with the amateur audience is larger than the average critic-amateur correlation (also see  Figure \ref{fig:correlationAudiences} left panel in the Supplementary Material). Second, the amateurs have greater dispersion in their observed taste similarities with other individuals. Thus, although the critics are quite representative of the amateur audience, it is not clear whether they are the best source of advice, and whether or how their ratings should be combined with those of amateur crowds.

\begin{figure*}[tbh!]
\begin{center}
\includegraphics[width=1\textwidth]{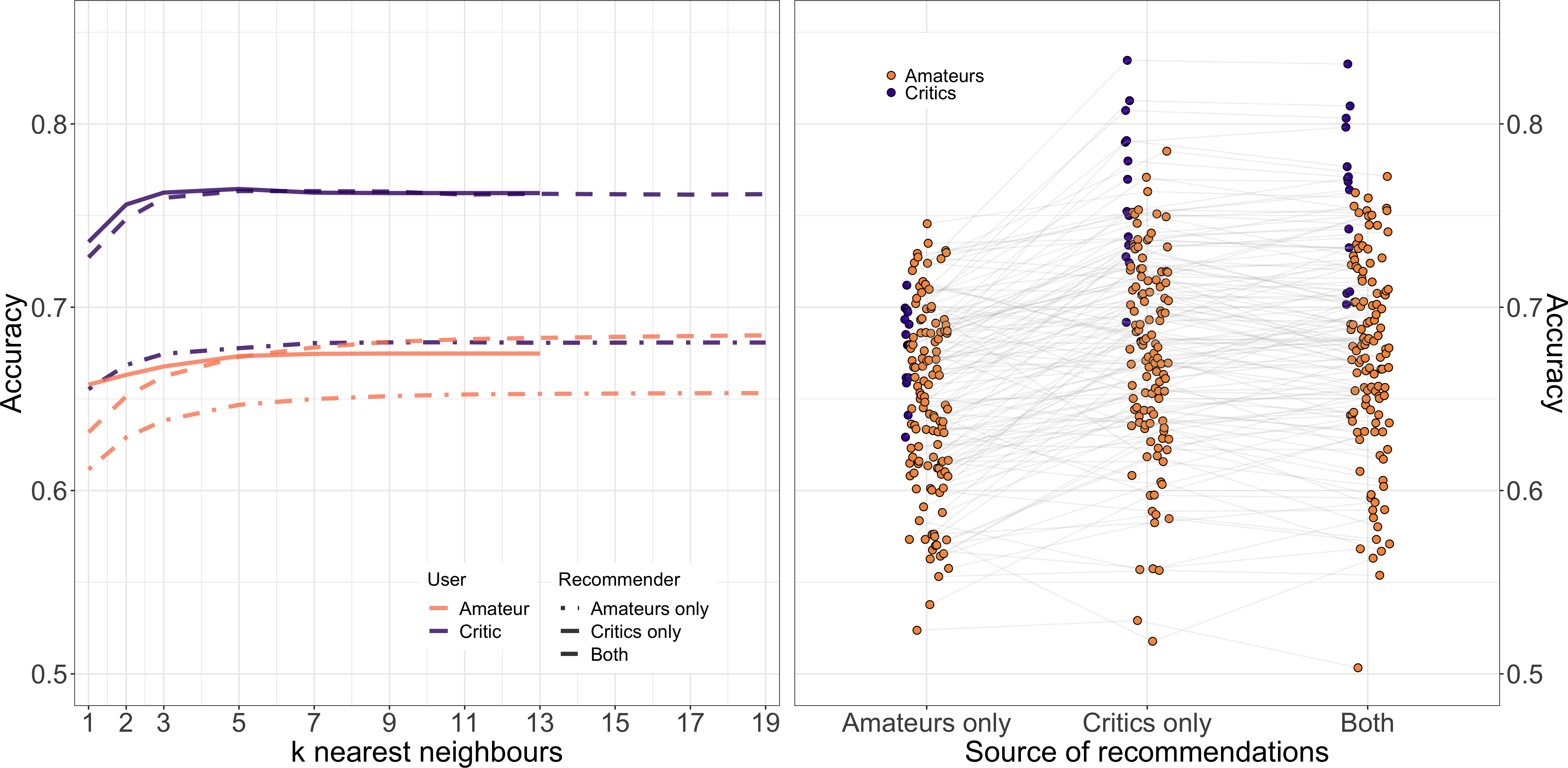}
\caption{\textbf{Performance of the recommender system for different groups (left) and individuals (right).}  \textbf{Left}: The average performance of the \emph{k-nn} algorithm based only on amateurs, only on critics, or both amateurs and critics for different values of $k$ for the amateur and critics groups. \textbf{Right}: The individual level performance of the \emph{k-nn} algorithm based only on amateurs, only on critics, or both amateurs and critics for $k=5$.}
\label{fig:amateursVsCritics}
\end{center}
\end{figure*}

\textbf{Following similar critics has high prediction value} We next compare the performance of recommender systems based on only amateurs or only critics in predicting the ratings of the amateur audience. When predicting amateurs, a recommender system based only on critics performs better than a recommender system based on amateurs --- the difference is substantial, and larger than 3\% in terms of prediction rate for the average amateur (compare the bold orange line with dash dot orange line in Figure \ref{fig:amateursVsCritics} left panel; the presented results are obtained by setting $\rho=1$) for any possible $k$ value. Even consulting the most similar critic can improve the performance by more than 1\% as compared to aggregating ratings from several similar amateurs (compare the leftmost point in the bold orange line with the rightmost point in the dash-dot orange line in Figure \ref{fig:amateursVsCritics} left panel). Taking advice from additional critics can marginally further improve performance. Because there is high consistency across critics,  there is not as much new information when considering the opinions of additional critics. These results hold regardless of the parameter $\rho$ used in the simulations (see Figure \ref{eq:amplification} in the Supplemental Material). In most cases, the observed performance differences translate directly to the individual level (see Figure \ref{fig:amateursVsCritics} right panel). That said, there is some variability in the population. For some amateurs, for example, consulting the one or two most similar critics is the best performing strategy, whereas for other individuals, taking advice only from amateurs performs best (i.e. see leftmost individuals in Figure \ref{fig:individualPerformance} in the Supplement and compare them to the individual at the bottom center).

\textbf{Ratings from critics and amateurs can complement each other} We next turn to the performance of a recommender system that uses the ratings of both critics and amateurs. For $k$ values lower than five, such a recommender system would perform modestly. In fact, people would be better off discarding the data from amateurs and considering only similar critics (compare the leftmost parts of the orange bold line and the orange dashed line in Figure \ref{fig:amateursVsCritics} left panel). The drop in performance for low $k$ values is substantial and further stresses the informational value of critics, as their ratings generalize better to unseen items than those of apparently similar amateurs. Still, for $k$ values equal to or larger than five, a recommender system using data from both critics and amateurs performs best for the amateur audience, even if only slightly so in our dataset, and can further improve predictive performance (see Figure \ref{fig:amateursVsCritics} left panel). This implies that the ratings generated from critics and amateurs can function as complements. Aggregating ratings from acclaimed critics and apparently similar amateurs in a single recommender system can help counteract the high statistical variance in the amateur ratings, and make the most of information from highly similar amateur individuals, leading to overall better predictive performance.

\begin{figure}[htbp!]
\begin{center}
\includegraphics[width=1\textwidth]{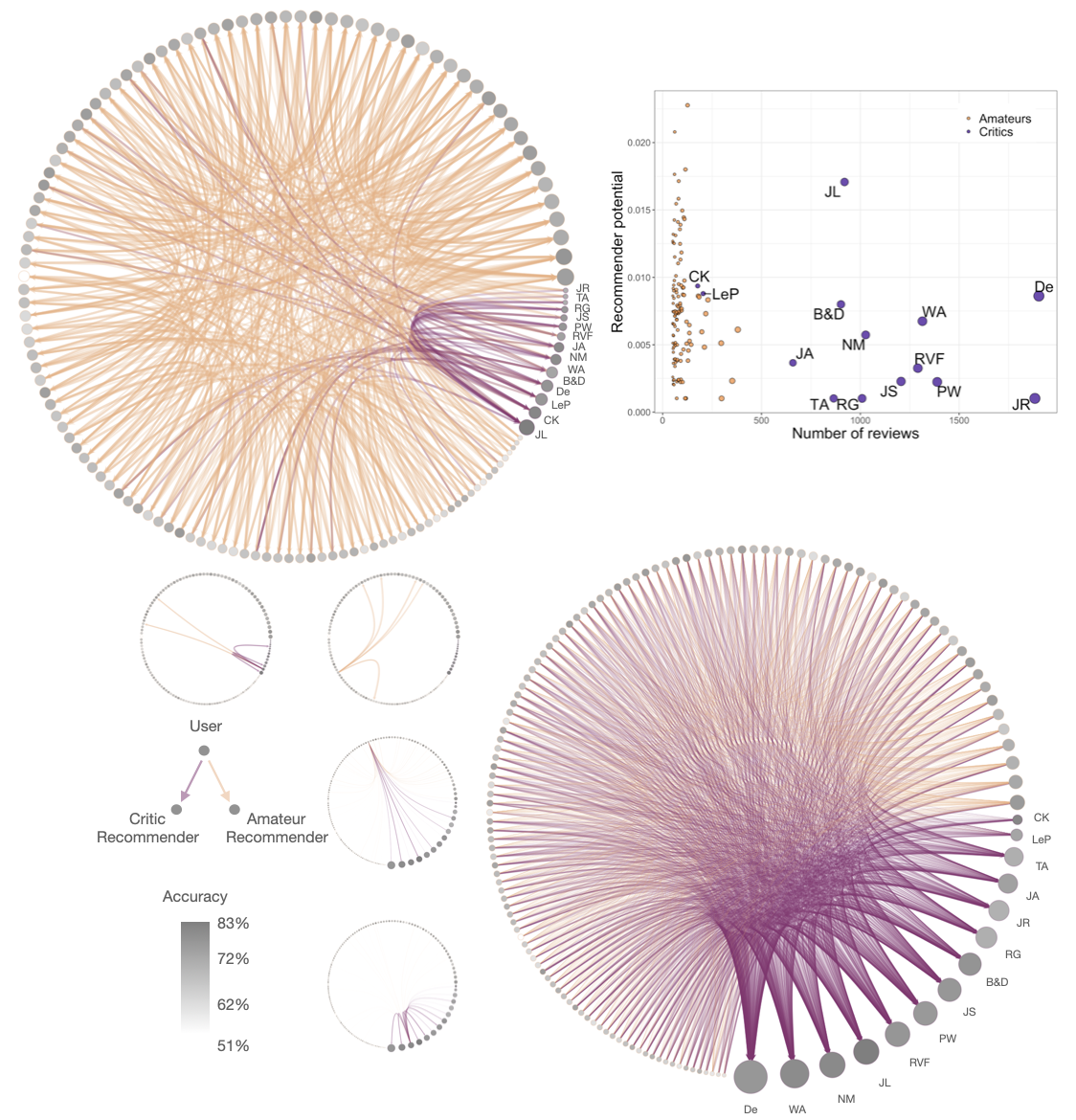}
\caption{\textbf{The recommender potential and recommender influence  of different individuals.}\\
Nodes represent individuals, node size represents recommender potential (upper left circle) or recommender influence (bottom right circle) of different people in the recommender network spanned by the \emph{k-nn} algorithm. \textbf{Bottom Left}: Orange edges indicate that advice is sought (or provided) from an amateur and purple edges indicate that advice is  sought (or provided)  from a professional critic. The colour of the nodes indicates the accuracy of the algorithm for different individuals in the dataset. Edges with weights smaller than 0.05 do not appear in the visualization to prevent overcrowding  the graph. \textbf{Upper Left}: The The advice-seeking network produced by the initial call of \emph{k-nn}, disregarding missing values (i.e., recommender potential). The edges (arrows) are pointing to the individuals from whom \emph{k-nn} first seeks advice for the target individual. \textbf{Lower Right}: The influence graph eventually produced by \emph{k-nn}. When an individual called by \emph{k-nn} has not rated a wine label, the next individual in the correlation rank is consulted. This process continues until $k$ advisers have been found or until the pool of potential advisers is exhausted. \textbf{Upper Right}:  Amateurs and professional critics placed on a 2-dimensional plane defined by the number of items they have evaluated (x-axis) and their recommender potential (y-axis). Critics are depicted with purple color and amateurs with yellow. Node size indicates the total influence of different individuals.}  
\label{fig:influencePotentialNetwork}
\end{center}
\end{figure}

\textbf{Critics (and critic-like amateurs) are more predictable} Regardless of the method used, critics are much more predictable than amateurs. Predicting critics' ratings using amateur ratings (the worst performing method for critics) leads, on average, to performance similar to predicting amateurs' ratings using ratings from both critics and amateurs (the best performing method for amateurs for high  $k$ values, see left panel in Figure \ref{fig:amateursVsCritics}). Further, when predicting critics' ratings the less-is-more effect persists at even higher $k$ values --- using data from amateurs and critics cannot improve performance compared to using data only from critics. In addition to being more consistent, critics are, on average, much more similar to everybody else (high mean-taste similarity) and they have rated many more wines. The latter two features can capture much of performance variability in collaborative filtering recommender systems \cite{adomavicius2005toward,analytis2023collaborative}. The same features can also predict recommender system performance when looking only in the amateur population. In fact, the data of several amateurs in our dataset can be predicted by the recommender algorithm with an accuracy similar to that obtained for critics (see the colour of the nodes in Figure \ref{fig:influencePotentialNetwork}A).  \footnote{A linear model using mean-taste similarity, dispersion in taste-similarity, and number of reviews as features could account for more than 65 \% of the variance in the prediction rates for different individuals for k = 5 and $\rho = 1 $ using data from both critics and amateurs (adjusted $R^2 = 
0.67$).} The exact same features are also predictive of people's potential to influence others in recommender systems \cite{analytis2020structure}, the topic to which we will turn next. 

\textbf{Some individuals have a larger potential to influence others} People differ substantially in their recommender potential (see Figure \ref{fig:influencePotentialNetwork}A), that is, the total weight they could have in recommendations made for all other individuals (i.e., the sum of advice weights attributed to them by \emph{k-nn}, not yet considering whether that individual actually rated a particular wine for which a recommendation is sought, also see Methods). Figure \ref{fig:influencePotentialNetwork}A, shows exactly how the \emph{k-nn} algorithm, seeks advice for the target individual from other individuals for $k = 5$ and $\rho = 1$. Among the critics, Jeff Leve has the largest recommender potential. Other world renowned critics such as Jancis Robinson and Tim Atkin score relatively low in terms of this metric. Further, a subset of the amateurs have larger recommender potential than most critics (see the points in the upper left side of Figure \ref{fig:influencePotentialNetwork}), which indicates that their rating patterns appear to be similar to those of many  other individuals in the amateur audience (also see Figure \ref{fig:correlationAudiences} left in the supplementary material). Some of these amateurs might have the potential to become influential critics.

\begin{figure}[htb]

\begin{center}
\includegraphics[width=1\textwidth]{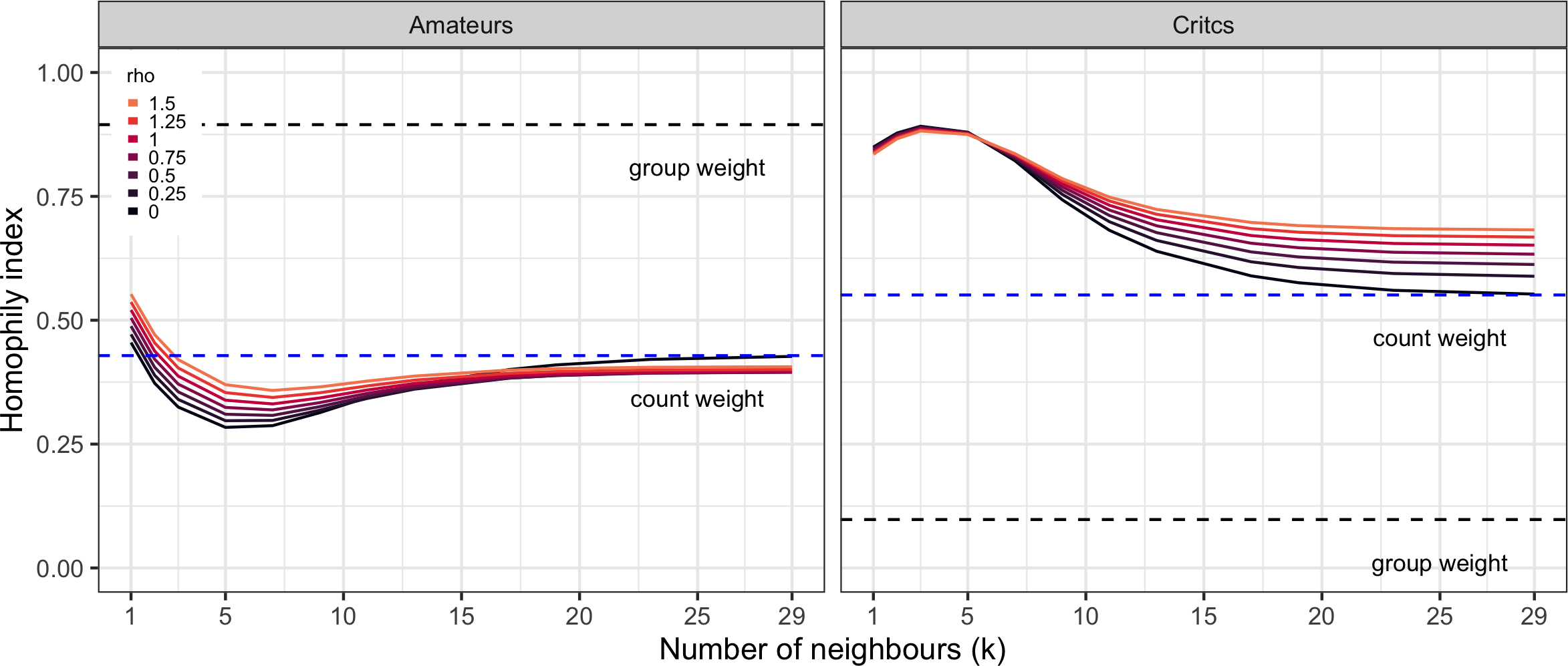}
\caption{\textbf{Homophily index of critics and amateurs} \textbf{Left and Right}: The homophily index of amateurs and critics as a function of the value $k$ in the $k$-nearest neighbors algorithm. Different $\rho$ values are represented with lines of different color. The horizontal dashed lines represent homophily baselines corresponding to the proportion of group members in the population (group weight) and the proportion of the ratings contributed by the members of each group (count weight).}  
\vspace{-1 em}
\label{fig:homophilyIndex}
\end{center}
\end{figure}

\textbf{The number of contributed ratings moderate people's influence} To derive estimates about how much people would enjoy specific wines, our implementation of the \emph{k-nn} algorithm calls in the adviser committee: the $k$ individuals with the highest correlations to the target individual who have evaluated a specific item. That is, for the same target individual, the adviser committee may differ from item to item depending on who has evaluated specific items. This also implies that recommender potential (defined by the initial choices of \emph{k-nn}, disregarding missing values) does not directly translate to recommender influence, as individuals with high potential may have evaluated only a few items. The lower right panel of Figure \ref{fig:influencePotentialNetwork} shows the recommender influence of different individuals in the population for $k = 5$ and $\rho  = 1$. On average, professional critics exert a much larger recommender influence than amateurs (5.54 vs .47) because they have evaluated many more items, and they are often consulted by the amateur audience in this recommender system. Even among the critics, however, there are some notable differences:  Jeff Leve, the critic with the highest recommender potential, recedes in recommender influence because he has evaluated only 40\% of the wines, while the journal Decanter, which has the largest number of rated items, would have the most influence in a wine recommender system built from these data. The recommender influence of different individuals can be also evaluated for specific wines, and depends on who else has rated that specific wine in the Supplementary Material (see Figure \ref{fig:influencePerWine}).

\textbf{The amateurs are (mostly) seeking advice from outside their group; the critics from inside} We next access the degree to which people get advice from individuals of the same or different group when the data from both groups were used in the recommender system. We calculated the homophily index (see Methods) of the amateur and critic groups for different values of $k$ and $\rho$ in the space of possible parameter configurations spanned by the $k$-nearest neighbors algorithm and we contrasted the index with the population proportion and rating proportion baselines (see Methods). When accounting for the proportion of ratings contributed from the two groups, the amateur group is characterized by slight inbreeding homophily, for low values of $k$, and heterophily for intermediate values of $k$ (see  Figure \ref{fig:homophilyIndex} left panel, the measure necessarily converges to the group's proportion of ratings for higher values of $k$). This result could partly explain the less-is-more effect in the recommender system performance for low values of $k$ because for these values advice is mostly drawn from apparently similar amateurs, resulting in high prediction variance. Higher $\rho$ values tend to increase homophily (decrease heterophily) for low $k$ values. Note that the amateur group is strongly heterophilous considering their proportion of the population because they draw less than 50\% of advice from other amateurs although they represent more than 85\% of the population. The wine critics, by contrast, are drawing most of their advice from other critics and are characterized by inbreeding homophily regardless of the baseline used. The homophily index in the critics group is particularly pronounced for a low number of neighbors $k$ and levels off as $k$ increases and more amateurs are (by necessity) consulted to advise the critics. For critics, higher $\rho$ values tend to lead to increased homophily for high $k$ values.

\section{Discussion}


Going back to the 18th century, the philosopher David Hume argued that critics are better equipped through training and natural predisposition to judge on matters of taste \cite{hume1757standard}, and his work has sparked philosophical debates about the potentially objective nature of judgement in matters of taste \cite{levinson2002hume}. Hume even preempted research on the wisdom of the crowds suggesting that two expert judges are better than one in matters of taste.  But are expert judgments on matters of taste representative of those of the wider public and should they be trusted? Further, do data from select critics complement or substitute for crowd-sourced data online or advice from friends? And is it possible to identify critics whose opinions are more informative or influential or even amateurs who have the potential to become successful critics? Last, how would information flow between critics and amateurs in a recommender system that relies on the \emph{k-nn} algorithm? We put forward a novel methodological framework that makes it possible to address these questions in any domain of taste, and applied it to the case of wine, a domain where expert judgement has been particularly revered.

\subsection{Studying expertise: from matters of truth to matters of taste}

Over the last few decades, disagreements among renowned wine critics such as Robert Parker and Jancis Robinson in regard to the quality of certain wine vintages have captured the attention of wine aficionados \cite{schatzker2016expert}. In addition,  empirical studies have raised doubt about the ability of expert wine judges to discriminate across wines in blind tastings \cite{hodgson2008examination}. Is there a good case for expertise in wine or in other taste domains? Scholars in applied psychology and the management sciences have identified a number of conditions that characterize expert judgment or prediction. Two key properties of expert judgment are (i) consistency across judges \cite{einhorn1974expert,mcauley2013amateurs}, and (ii) or discriminating ability \cite{shanteau2002performance} (i.e. or the ability of judges to evaluate similar items with similar ratings). Previous work on wine experts in blind tasting settings has noted average correlation among them ranging from 0.2 to 0.5 depending on the study \cite{ashton2012reliability,cicchetti2004won,brien1987analysis}. In our data, we clearly see that renowned wine critics are highly consistent---the average mean correlation between each individual critic and other critics is higher than 0.5 for all individual critics. Thus, the average correlation rate is higher than reported in previous studies on red wines and similar to what is observed in some domains of facts in medical and business contexts \cite{ashton2012reliability}. The increased level of consistency across judges might reflect the higher level of expertise of the critics included in our study, who are arguably some of the most renowned tasters in the world, and might share the same evaluative schemas or the fact that en Primeur ratings are intended to be predictive of future wine quality and are not blind. Our results also imply that critics are more discriminatory than amateurs---the levels of observed agreement would not be possible if their judgments were not guided by clear and shared criteria about what makes a good wine or if their judgments were too noisy \cite{broomell2009experts}. Overall, our findings indicate a high level of agreement among critics and show that that expert judgments for wine are characterised by similar statistical properties as in some domains of truth.

\subsection{Amateurs and critics' ratings: substitutes or complements?}
In the past, a number of studies have looked at correlations between critics and amateur raters. In their majority, these studies have found that amateur and expert opinions are only mildly correlated \cite{holbrook1999popular,holbrook2005role,wanderer1970defense}. Although valuable, studies comparing the tastes of different groups using correlations leave a lot to be desired---correlations do not immediately translate to predictive power, especially when the opinions of several judges are aggregated. Going beyond previous work we evaluated the performance of different strategies that people can use to get recommendations or predict their own tastes, when seeking advice from critics, amateurs or both groups. We found that relying on the opinions of just one critic led to better predictive performance for most amateurs than seeking advice from several other amateurs. This is because critics are both more consistent and prolific in their evaluations. Having evaluated many wines helps to correctly estimate correlations between amateurs and critics and to effectively generalize in unseen data. Hume's conjecture that two critics are better than one was vindicated in our analyses, but the additional gains are modest. Overall, in the case of wine, high observed similarities with critics tend to be robust, and therefore the ratings of critics can be valuable proxies that help people predict what they like. This may explain why many renowned critics have been able to monetize the information value of their reviews by introducing subscriptions to their websites, and why the \emph{following-the-most-similar critic} heuristic is a common decision strategy among wine afficionados \cite{taber2006judgment}. 

There is one previous study from the recommender systems community that allows us to compare our results with those from another domain: Amatriain and colleagues compared a \emph{k-nn} recommender system based on the ratings of select film critics to a system based on the ratings of thousands of amateurs \cite{amatriain2009wisdom}. They found that recommender system users receive slightly more predictive recommendations when relying on a large database with opinions from other amateurs than when drawing recommendations from a database of select critics (but using movie critics has a number of advantages in terms of scalability, privacy, etc.). The somewhat diverging results could reflect fundamental differences in the acquisition of expertise in the two domains: whereas watching films is almost equally accessible to everybody, and is mostly constrained by the time budget that people have available, opening and tasting wines is mostly subject to budget constraints, and can quickly become an expensive hobby, reducing the number of training samples that most amateurs have access to, but also their capacity to supply information to others. This is also reflected in the relative volume of ratings supplied by critics as compared to amateurs, which is larger in our study. 

Last, going beyond considering information generated from critics and amateurs separately, as in the study by Amatriain and colleagues, we also examined what happens when information is drawn from both critics and amateurs, and found that for large values of $k$ there is a margin to further improve recommendations. This indicates that the data points generated by critics and the potentially more voluminous data generated by amateurs in online interfaces could be complementary in a collaborative filtering recommender system. What is more, our methods make it possible to identify the informational contributions of different individuals or groups for specific items (see Figure \ref{fig:influencePerWine} in the Supplement) and in the aggregate.

\subsection{From real-world to in silico advice networks and back}

Following the groundbreaking work of Katz and Lazarsleld in the 1950s \cite{katz2017personal}, social scientists have used survey methodologies to elicit advice networks across domains of life (including fashion, politics, and beyond) as well as to uncover the structure of professional advice networks \cite{lazega2012norms} and informal organizational networks \cite{krackhardt1987cognitive}. One of the main findings emerging from this research stream is the existence of informal opinion leaders who are sought for their advice by many other individuals. Similar to the opinion leaders of real-world advice networks, some individuals are much more often sought by the \emph{k-nn} algorithm and provide advice to many similar others \cite{analytis2020structure}. Our analysis routines allowed us to uncover the position of critics and amateurs in such in wine advice networks in silico and showed how their influence is modulated by the parameters $k$ and $\rho$ of the \emph{k-nn} algorithm, but also by the statistical properties of people's tastes and whether they are prolific raters. As such, the networks produced by \emph{k-nn} can be seen as informationally efficient advice networks for matters of taste and can be compared in terms of their structure and properties to the advice networks formed by people offline or in online platforms \cite{lewis2008tastes,lewis2018conversion}. Thus, the networks produced by \emph{k-nn} could provide new hypotheses about the formation of real-world advice networks and can help disentangle informational and other motivations when forming new ties with potential advisors.



\subsection{Relative expertise and identifying talent}

In domains of fact it is relatively easy to identify who is the best or the most accurate judge by looking at the prediction success of different judges in past data \cite{shanteau2002performance,kurvers2016boosting}. Is there a way to assess the relative worth of judges in matters of taste, where there is no objective truth to be predicted or when long records of past data are not available? This would allow us to identify critics whose opinions are particularly valuable to others, as well as to identify talented amateurs who have the potential to become critics. One possibility is to use the average ratings for an item (i.e. wine) as a gold standard and to assess how different judges can predict it \cite{carayol2019evaluating}. Although this could be a good assumption in some settings, it does not do full justice to the subjective nature of tastes and could break down in domains where tastes are polarized. Another approach would be to use correlations with other individuals (judges or raters) as a proxy for informational quality \cite{kopsacheilis2023crowdsourcing,kurvers2019detect,atanasov2023talent}. This strategy has already shown some promise in settings where there is an objective correct answer---Kurvers et al. \cite{kurvers2019detect}, for example, have shown that using the average correlation of an individual with other judges is a reliable heuristic for judge quality in several domains, and it can lead to good predictive performance when the truth cannot be immediately verified. The approach we put forward in this paper is similar in spirit and generalizes the similarity principle in matters of taste: it relies on the node strength resulting from the advice network produced from the \emph{k-nn} algorithm to identify influential critics and talented amateurs. Using this method we were able to identify Jeff Leve as the critic with the highest recommender potential, but also to spot several amateurs who appear to have a large capacity to inform others. Our method could be used by online platforms such as Vivino to identify and nurture talent in its user base.  


\subsection{Homophily and polarization in matters of taste}
Many real world opinion spaces are polarized. People tend to interact, listen to and get influenced by other individuals belonging to the same groups, a property commonly referred to as homophily. The tendency to interact with similar people might be further reinforced by personalization algorithms we use in our every-day life, such as the \emph{k-nn} algorithm that we investigated in this paper. In a similar vein, it has been argued that some personalization technologies may lead people into filter-bubbles, where most of the information they consume comes from similar individuals \cite{pariser2011filter}. In the case of tastes, for instance, sociologists have long argued that people belonging to different classes, cultures, or even political affiliations also differ in their aesthetic preferences  \cite{veblen2005theory,park2017cultural,bourdieu2012distinction,mcpherson1987homophily}. Therefore, one would expect that at least in some domains of taste collaborative filtering algorithms might produce homophily and taste filter bubbles. 

We developed a method that can be used to study whether the influence networks generated by the \emph{k-nn} algorithm produce informational insulation for different groups by adapting a well established metric of homophily and applying it to the domain of taste.  When accounting for the number of ratings contributed by the two groups (count weight), the amateur tastes would be characterized by slight homophily for low values of $k$, and by slight heterophily otherwise. By contrast, critics would get most of their information from other critics regardless of how the \emph{k-nn} algorithm is configured. The methods that we developed in this study can be readily applied to study taste homophily in collaborative filtering systems in any other taste domain and for other types of categorical groups (e.g.  men and women, Europeans and Americans, etc.). For example, Dellaposta and colleagues \cite{dellaposta2015liberals} recently pointed out that people's differences in political convictions are also reflected in their tastes. Thus, using our methods one could test whether a collaborative filtering system would generate recommendations for target users by drawing information from people with similar political affiliations, potentially further increasing the cultural divide between different groups or people.

\subsection{Conclusion}

  In vino veritas---in wine, there is truth---says an old Latin adage. We found that critics' judgements are indeed valuable in helping amateurs identify good wines, more so than the opinion of most other amateurs. Still, there is scope for combining the opinions of both critics and amateurs, and for identifying the most influential critics and talented amateurs, whose tastes appear to be informative for many other individuals. Going beyond wine, the methods we developed are modular and generic and can be readily applied in any dataset where different groups of people have rated a number of items, even when there are discrepancies in the number of items evaluated, to tackle long-lasting research questions in the social, behavioral and management sciences. Further, they can be also used as a tool for improving people's understanding of  key collaborative filtering algorithms \cite{verbert2013visualizing} by enhancing the transparency and interpretability of the recommendation process \cite{herlocker2000explaining}, and can help people hone in on the right decision strategies when seeking a good item in matters of taste.

\section*{Competing interests}
  The authors declare that they have no competing interests.

\section*{Acknowledgements}
We would like to thank Orestis Kopsacheilis, Sune Lehmann, and the members of the Strategic Organization Design and the Culture, Consumption and Commerce groups at the University of Southern Denmark for helpful discussions. Pantelis P. Analytis is supported by a Sapare Aude grant by the Independent Research Fund Denmark.  Bahador Bahrami is supported by the European Research Council (ERC) under the European Union’s Horizon 2020 research and innovation programme (819040 - acronym: rid-O), and by the Templeton Religion Trust. Ophelia Deroy is supported by the Momentum grant co-Sense from Volkswagen foundation, and the BIDT Co-learn grant. We thank Rhett Nichols for editing this manuscript.

\section*{Code availability statement}

Our code and datasets are available at https://osf.io/pqaw3/. 

\bibliographystyle{plain}
\bibliography{criticsAndCrowds}

\newpage

\section{Supplementary Material}
\subsection{Correlation profiles in sparsity balanced data}

In the results we reported in the main text the number of ratings in the two groups was unbalanced because the critics have evaluated many more of the items than the amateurs. This imbalance is an attribute of the wine market in general (i.e. critics are much more prolific in ratings than amateurs). To assess whether the correlational results we observed are influenced by this imbalance we artificially reduced the ratings of critics by removing ratings from each critic at random until their data density was approximately equal to that of the average amateur ($4.7 \%$). We then repeated the correlation analysis reported in the main text by calculating the within group correlations and the correlations with the entire population. Because the results from a single random sample of this reduced dataset can be very noisy we repeated the process 1000 times and averaged the results across repetitions. The average mean taste similarity and the dispersion in taste similarity for the critics has largely remained unchanged as compared to the results presented in Figure \ref{fig:winePlane}. Thus, although density has a direct impact on the recommender influence of different individuals it does not affect their expected correlations or their recommender potential substantially.  

\begin{figure}[htbp!]
\includegraphics[width=1\textwidth]{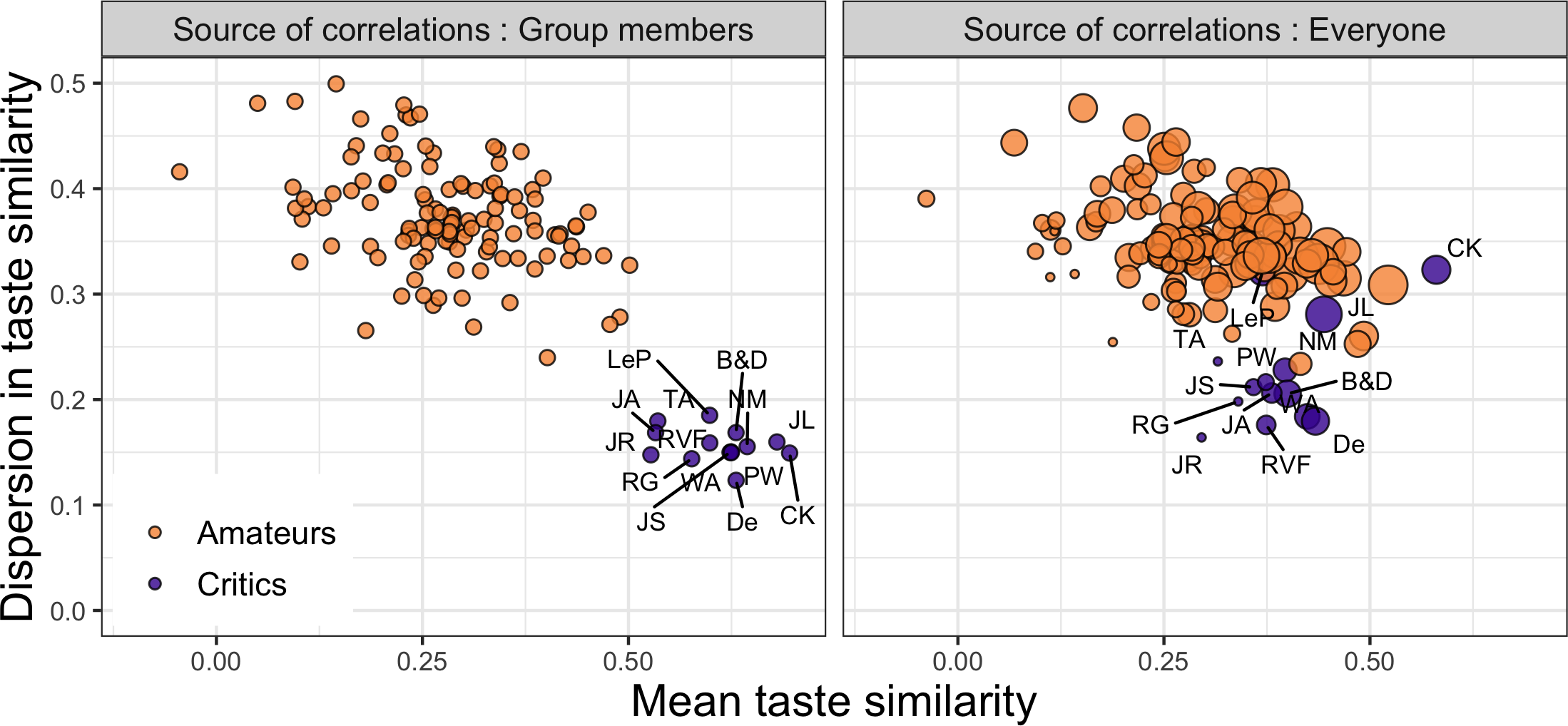}
\vspace{-7mm}
\caption{\textbf{Intercorrelations with individuals belonging in the same group and all other individuals when the data are balanced for sparsity.}  \textbf{Left}: The position of 14 professional critics and 120 amateurs on a 2-dimensional plane defined by mean taste similarity (i.e., mean correlation) and dispersion in taste similarity (i.e., standard deviation of correlations) with members of the same group. \textbf{Right}: The position of the same 14 professional critics and 120 amateurs on the same plane, but this time with taste similarity calculated across all individuals (professional critics and amateurs). The color in both panels indicates whether an individual is a professional critic or an amateur and the point size in the right panel indicates the recommender potential of different individuals for $k = 5$ and $\rho=1$.}  
\label{fig:correlationDensitySparse}
\end{figure}


\newpage
\subsection{Similarity to the amateur and critic audiences}

\begin{figure}[htbp!]
\vspace{-2mm}
\includegraphics[width=0.9\textwidth]{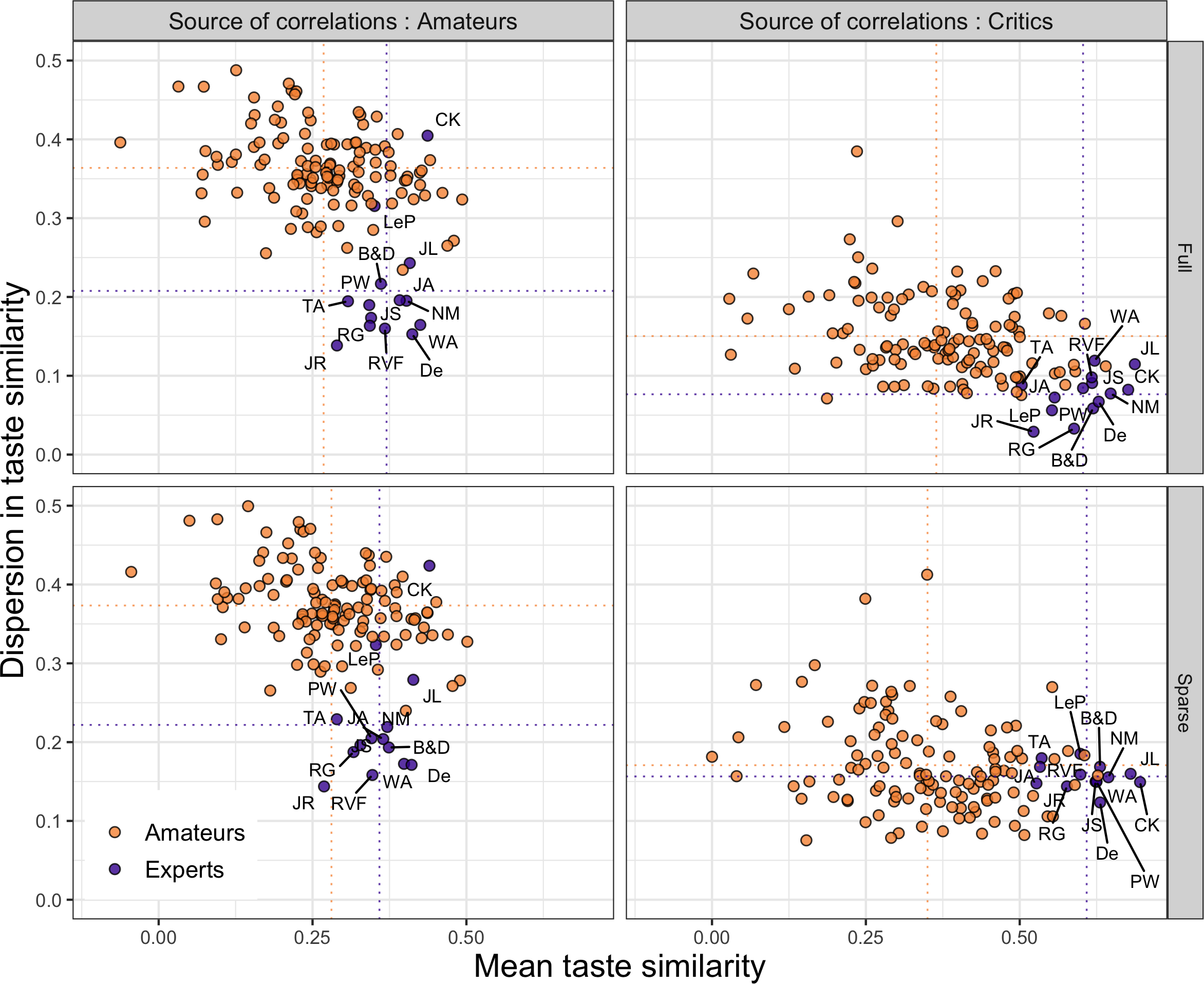}
\vspace{-4mm}
\caption{\textbf{Intercorrelations with amateurs and critics}  \textbf{Left}: The position of 14 professional critics and 120 amateurs on a 2-dimensional plane defined by mean taste similarity (i.e., mean correlation) and dispersion in taste similarity (i.e., standard deviation of correlations) with amateurs. \textbf{Right}: The position of the same 14 professional critics and 120 amateurs on the same plane, but this time with taste similarity calculated across all individuals (professional critics and amateurs). The color in both panels indicates whether the an individual is a professional critic or an amateur. In the bottom row we repeat this analysis but for data balanced for sparsity. The dotted orange and purple lines indicate the average correlations and dispersion in taste similarity recorded for that group of people.}  
\label{fig:correlationAudiences}
\end{figure}


In this section we look at people's similarity exclusively with critics or amateurs, both for the full dataset and also when balancing the data for sparsity (as in the previous section). Figure \ref{fig:correlationAudiences} top left shows the correlations of the two groups with amateurs. It reveals that every single critic has correlations with amateurs higher than the average correlation among the amateurs themselves (all purple dots are on the right on the vertical dotted orange line). Figure \ref{fig:correlationAudiences} top right shows the correlations of the two groups with critics. The graph reveals that critics are more correlated among themselves than amateurs are with critics. Nonetheless,  some amateurs are similar to critics in their correlation profiles, and it would have been hard to distinguish them from critics if we did not impose the critic/amateur categorization (compare with Figure \ref{fig:winePlane} left, where a separation of the groups would be possible merely using their correlation profiles). These results change very little when we remove ratings from the critics until their data density becomes similar with the average amateur density (Figure \ref{fig:correlationAudiences} bottom left and right).

\enlargethispage{20mm}

 \newpage
\subsection{Strategy performance as a function of the weighting scheme}

The average performance of the weighted k-nearest neighbor algorithm clearly depends of $k$, with larger $k$ values leading to better performance. But how does the weighting scheme, as expressed by the parameter $\rho$ alter the performance of the different recommendation approaches? To explore this question we show how performance varies as a function of $\rho$ for different $k$ values. Remember that $\rho = 0$ implies equal weights, while when $\rho = 1$ the weights directly correspond to the correlations. Overall, there appears to be little performance variation as a function of $\rho$ (less that 2 \% for our entire parameter space). For critics, larger values of $\rho$ lead to slightly better performance, especially for high $k$ values and irrespective of the source the data is drawn from. For amateurs, higher $\rho$ works best when drawing advice from critics, equal weighting ($\rho = 0$) works best when drawing advice from amateurs, and intermediate $\rho$ values work best when drawing advice from both groups (with the exception of low $k$ values, e.g. $k=3$, where equal weighting leads to the best results).

\begin{figure}[htbp!]
\includegraphics[width=1\textwidth]{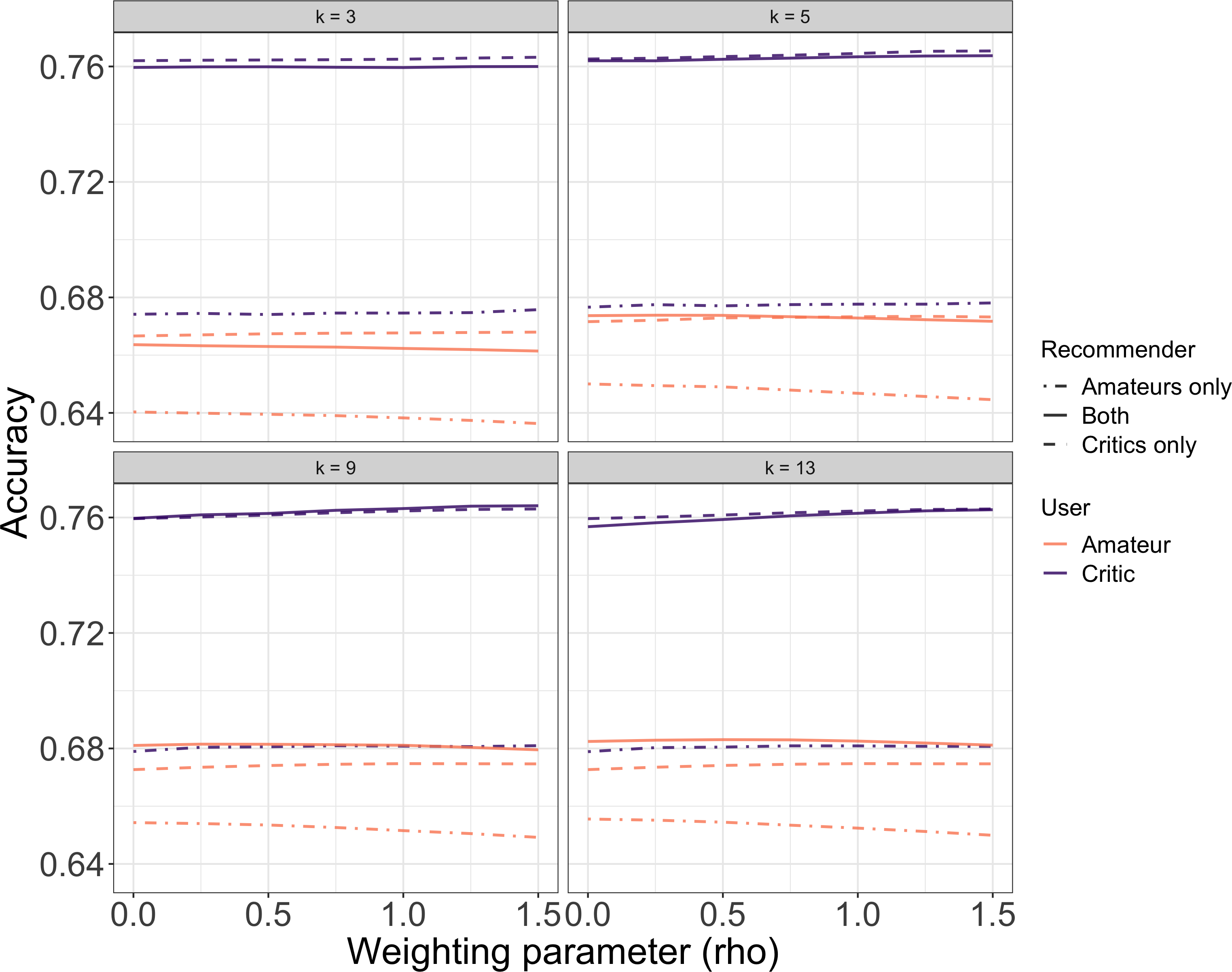}
\vspace{-7mm}
\caption{\textbf{Performance of the \emph{k-nn} algorithm as a function of the parameter $\rho$ for different  $k$ values}  \textbf{Each panel}: The average performance of the \emph{k-nn} algorithm based only on amateurs, only on critics, or both amateurs and critics for different values of $k$ (different panels) for the amateur and critics groups, while varying the parameter $\rho$ of the algorithm (x-axis).}  
\label{fig:weightSkewness}
\end{figure}

\enlargethispage{30mm}

\newpage

\subsection{Influence networks for individual wines}

In this section we present examples of the influence networks for specific wines with varying popularity. For each single wine, for each repetition of the simulation, and for each target individual, the k-nearest-neighbors algorithm is searching for the $k$ most similar individuals in the population who have evaluated the wine. Note that in each run of the simulation, and for each individual 10 items are withheld uniformly at random. This implies that some advisers are not available in some of the 1000 runs of the simulation and there is slight variation in the people who eventually can provide recommendations. By comparison, the influence network presented in Figure \ref{fig:influencePotentialNetwork} in the main text is constructed by aggregating the influence networks of every single wine label included in our collection.

\begin{figure}[htbp!]
\begin{center}
\begin{tabular}{cc}
  \includegraphics[width=.35\textwidth]{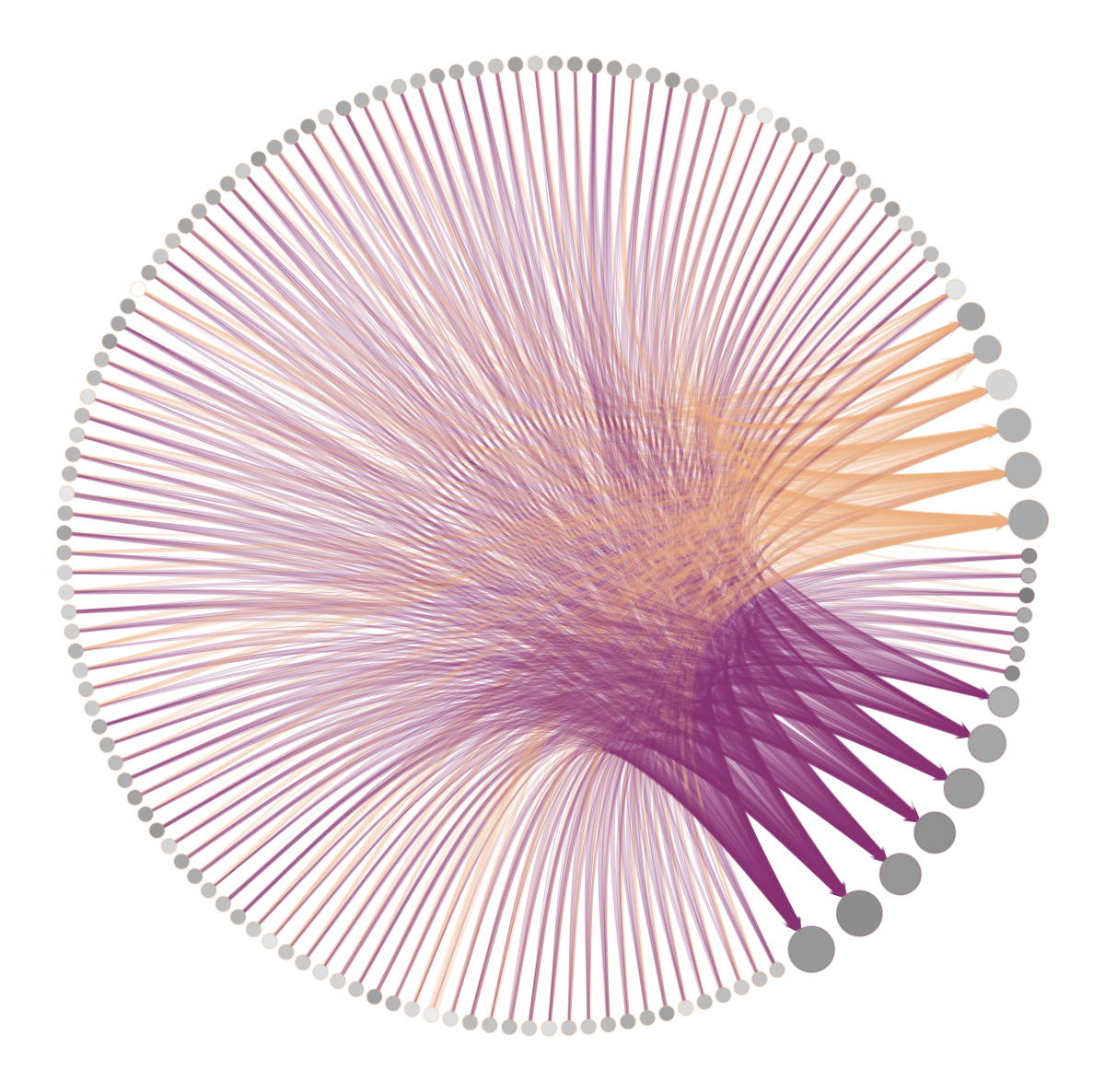} &   \includegraphics[width=.35\textwidth]{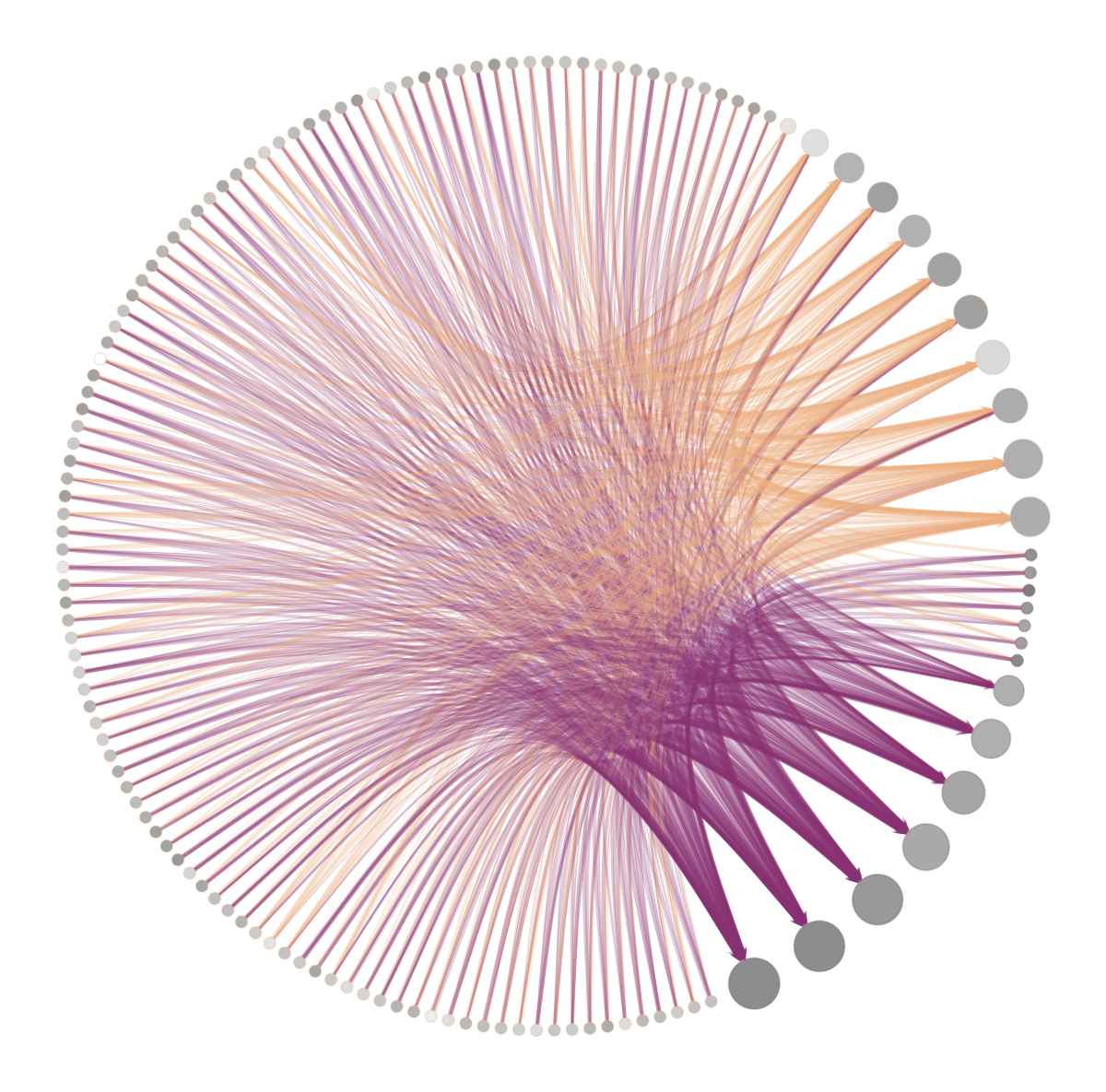} \\
(a) du Tertre 2004 & (b) Bernadotte 2007 \\[6pt]
 \includegraphics[width=.35\textwidth]{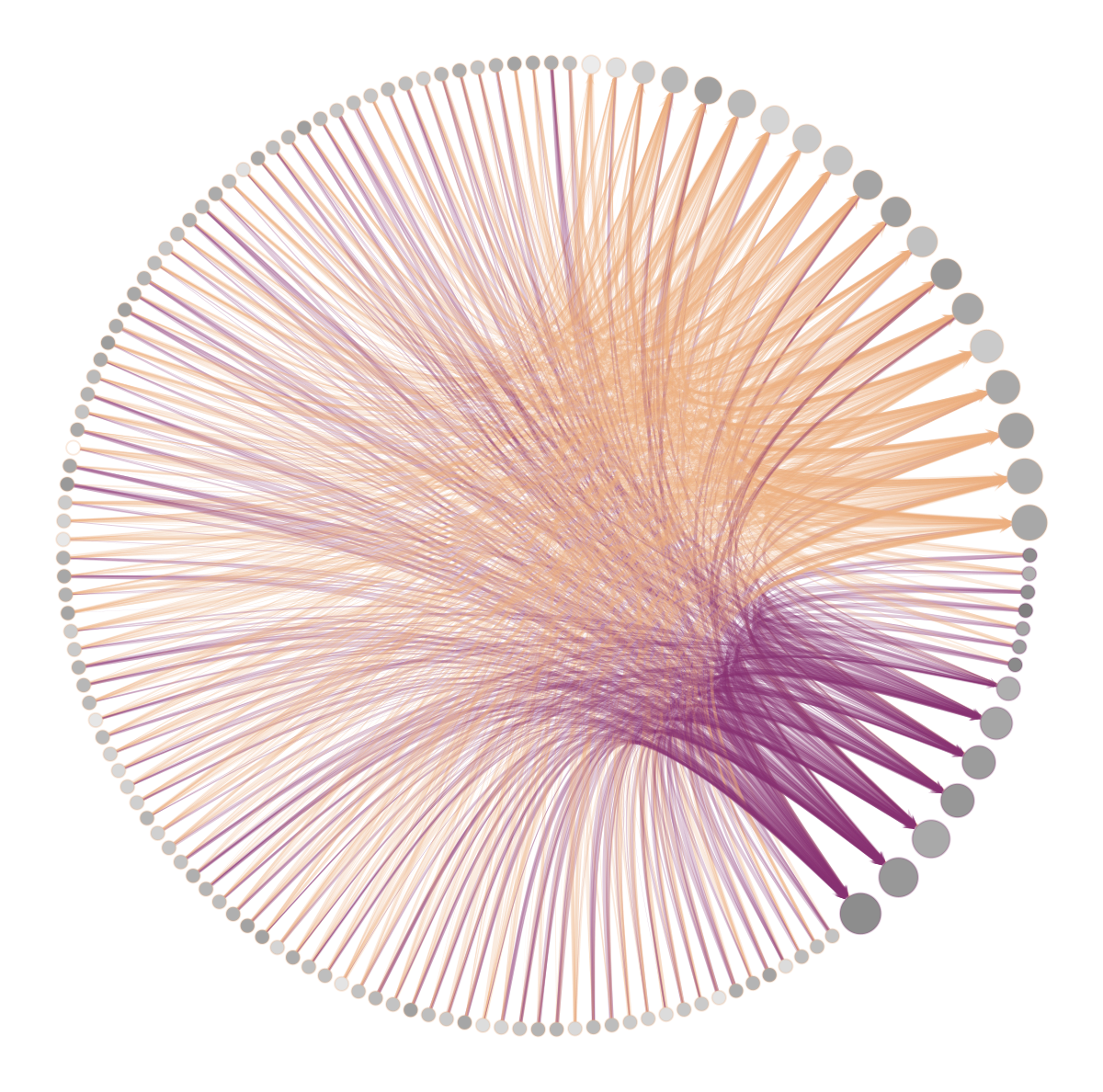} &   \includegraphics[width=.35\textwidth]{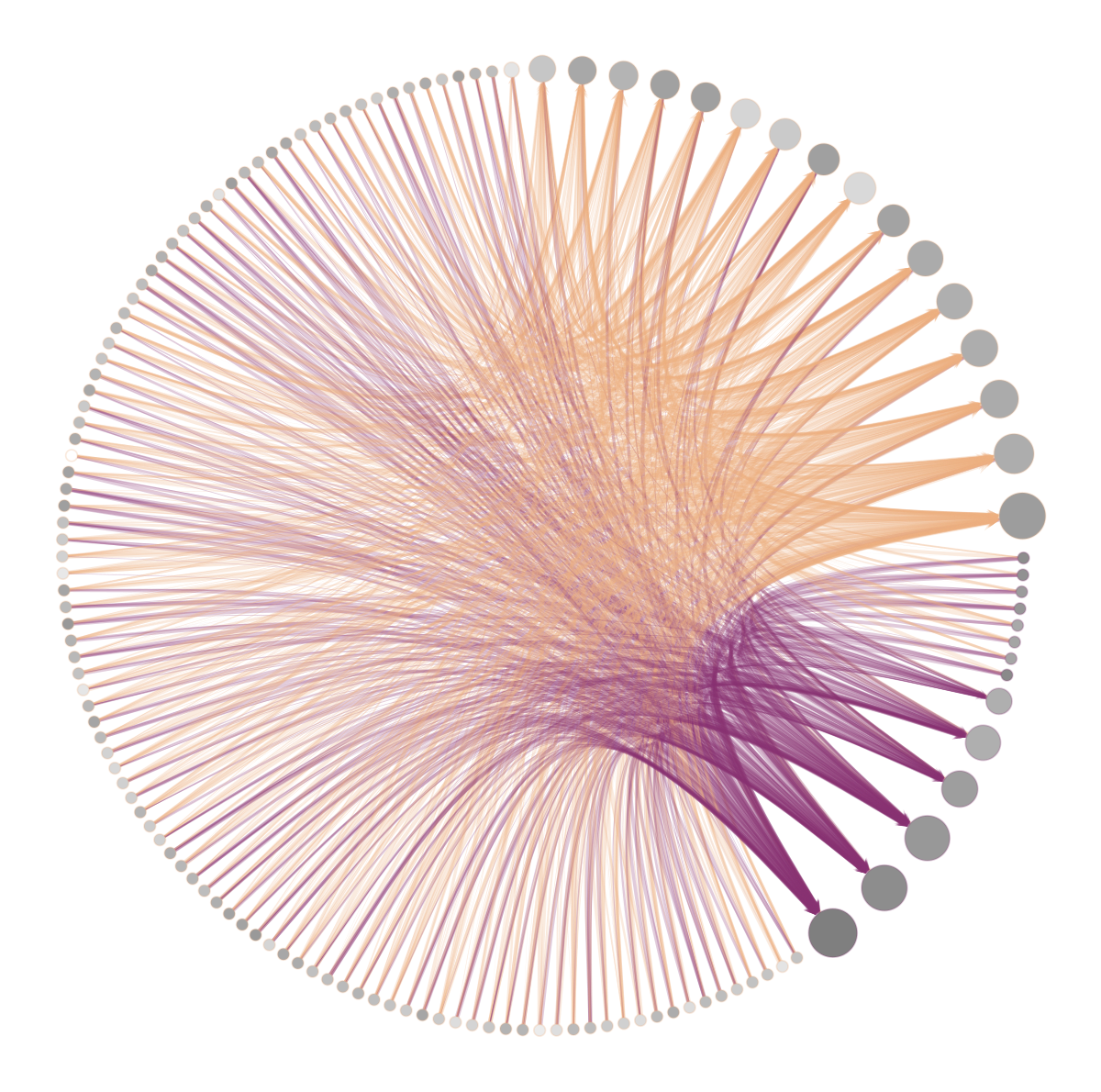} \\
(c) Pontet Canet 2008 & (d) Leoville barton 2013 \\ [6pt]
\end{tabular}
\caption{\textbf{The influence network for specific wine labels with varying popularity} Nodes represent individuals and the size of the nodes represents the recommender influence of different individuals in the recommender network spanned by the k-nn algorithm. Orange edges indicate that advice is sought (or provided) from an amateur and purple edges indicate that advice is  sought (or provided)  from a professional critic. The color of the nodes themselves (from light to dark grey) indicates the estimation error of the algorithm for different individuals in the entire dataset. Edges with weights smaller than 0.05 do not appear in the visualization to prevent overcrowding  the graph.}
\label{fig:influencePerWine}
\end{center}
\end{figure}

\newpage

\subsection{Homophily index using only the initial calls}

\begin{figure}[htb]

\begin{center}
\includegraphics[width=1\textwidth]{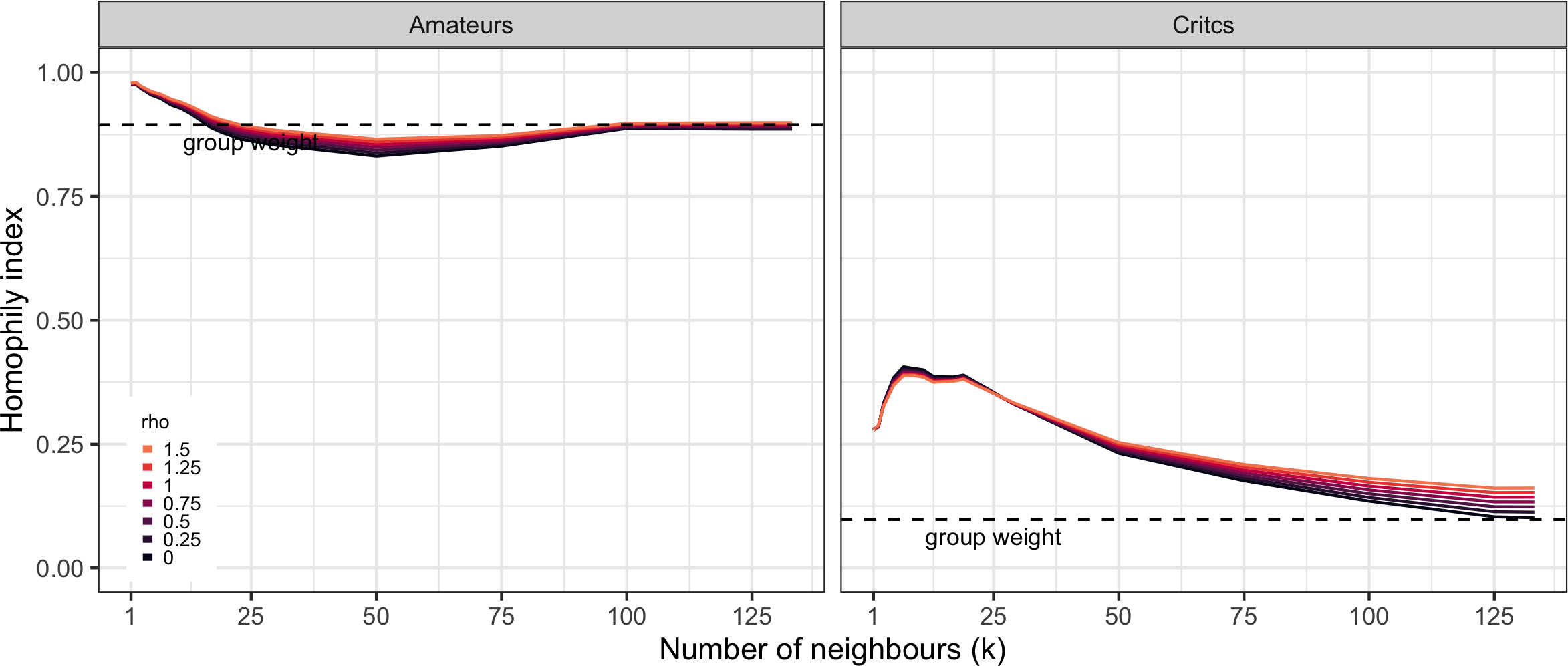}
\caption{\textbf{Homophily index of critics and amateurs}: The homophily index of amateurs and critics as a function of $k$ in the \emph{k-nn} algorithm when we consider only the first calls of the algorithm. Different $\rho$ values are represented with lines of different color. The horizontal dashed line represents a baseline corresponding to the proportion of group members in the population (group size).}  
\vspace{-1 em}
\label{fig:homophilyIndexPotential}
\end{center}
\end{figure}

The homophily index can also be seen as a measure of preference (or bias) for obtaining information from members of the same group. In the main text we presented results in terms of the ratings actually used by the \emph{k-nn} algorithm to inform people's choices. However, when using actual ratings, the strength of the preference is not fully expressed because missing ratings from amateurs could be often substituted by the ratings of critics (who tend to be more prolific). Thus, we also visualize the homophily index when only the first $k$ individuals called by the algorithm are included, and disregarding whether people eventually contributed ratings. This measure of homophily gives a more direct impression of the algorithm's predilection for using information from members of the same group. This measure replicates the gist of the results presented in the main text. The amateurs are characterized by inbreeding homophily for low to intermediate values of k (<17) and become slightly heterophilous for values above that.  The critics, by contrast, are characterized by substantial in-breeding homophily for low values of $k$. As $k$ increases, and more people are sought for advice, the homophily index  converges to the group weight (for $\rho = 0$ it converges exactly to the group weight whereas for higher $\rho$ values  there might be small deviations due to the weights assigned to different individuals, see the right panel of Figure \ref{fig:homophilyIndexPotential}.)

\newpage

\subsection{From aggregate to individual level performance}

\begin{figure*}[tbh!]
\vspace{-0.6 em}
\begin{center}
\includegraphics[width=0.9\textwidth]{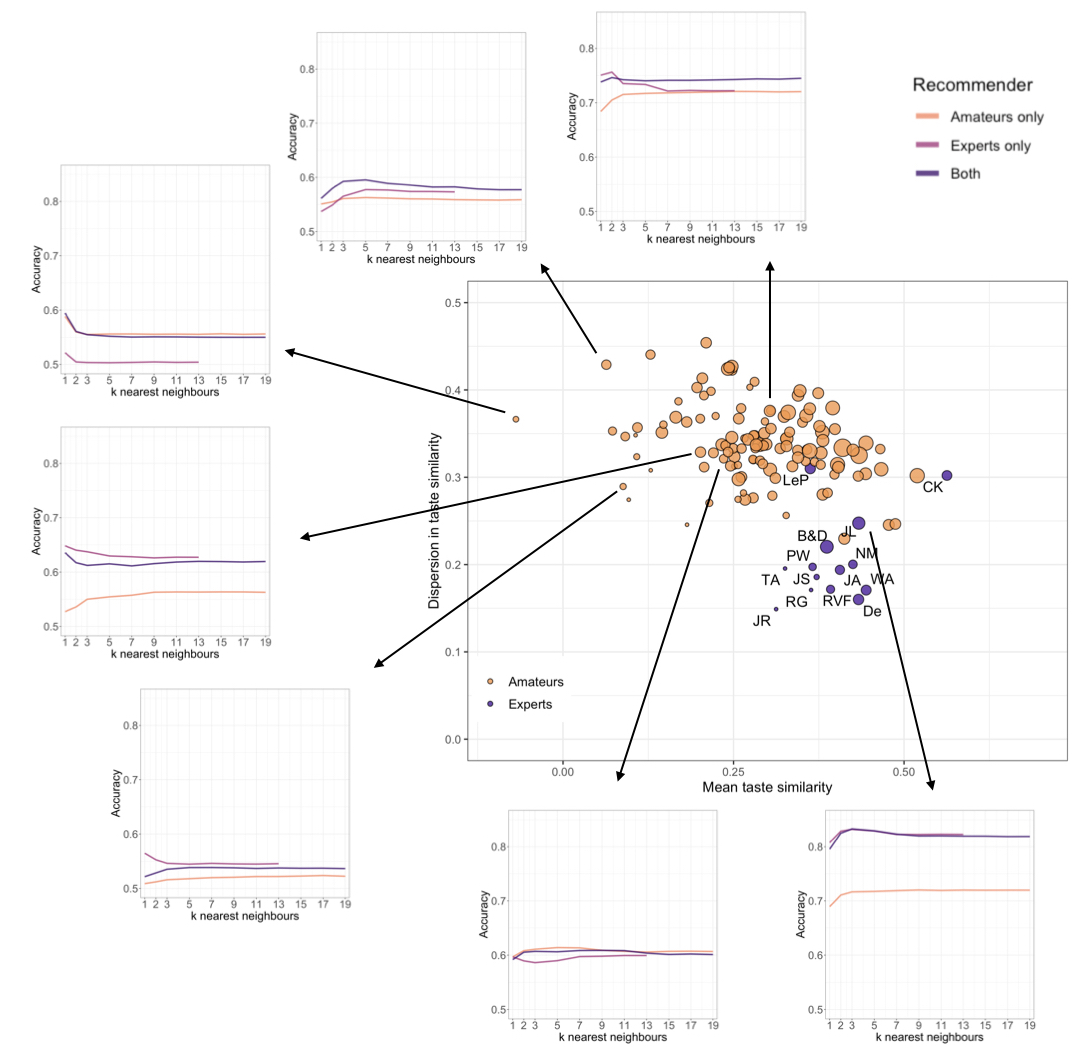}
\caption{\textbf{The position of the 14 professional critics and 120 amateurs on the 2-dimensional plane defined by mean taste similarity and dispersion in taste similarity, but this time with taste similarity calculated. The color in both panels indicates whether an individual is a professional critic or an amateur and the point size indicates recommender potential.}
}  
\label{fig:individualPerformance}
\end{center}
\vspace{-2.2em}
\end{figure*}

\enlargethispage{20mm}

So far we have presented how the performance of different algorithms varies as a function of $k$ and $\rho$ when we average across individuals. This analysis can be also be repeated at the individual level to provide more nuanced results on how the performance of the \emph{k-nn} algorithm (or different social learning strategies) changes for different individuals and as a function of the number of neighbors $k$ for different values $\rho$. In Figure \ref{fig:individualPerformance} we demonstrate the power of such individual level analysis for $\rho$ = 1. We present three individuals for whom the heuristic \emph{follow-the-most-similar-critic} was the best or a nearly the best strategy to follow (middle and bottom left and top right), one individual for whom following the most similar amateur performed best (top left), one individual for whom the best solution was following a clique of amateurs (bottom center), and one for whom the opinions of critics and amateurs where clearly complementary (top center). Last, we present the performance of the algorithm for Jeff Leve, the critic with the highest recommender potential (bottom right).

\newpage

\subsection{Correlation profiles of critics}

\begin{figure*}[tbh!]
\begin{center}
\includegraphics[width=.65\textwidth]{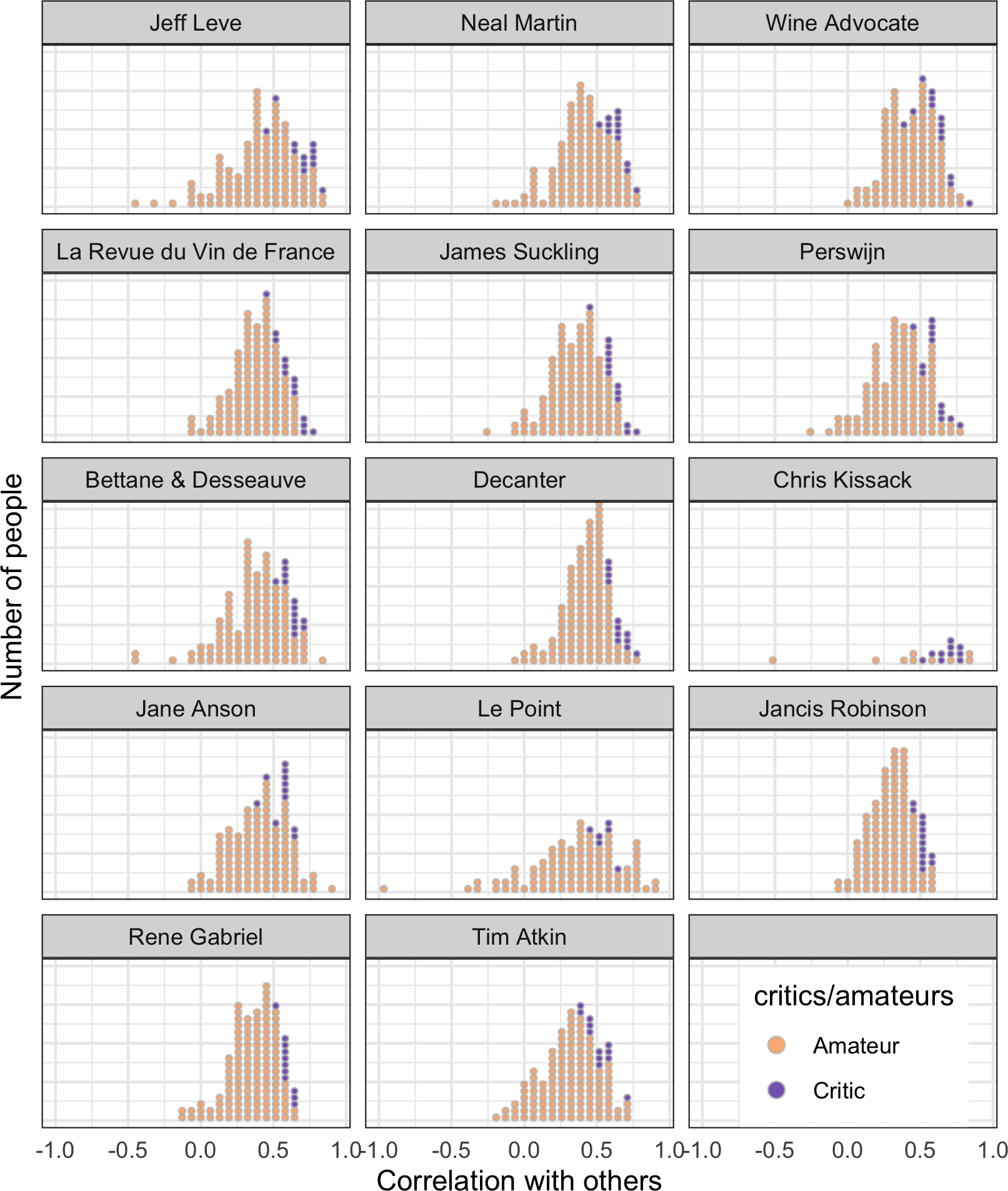}

\caption{\textbf{The observed correlations of the 14 critics with other critics  and amateurs. Orange beads correspond to correlations with amateurs and purple ones to their correlations with critics. We do not report correlations where the critic had less than 5 items in common with other individuals. The critics have been ranked in descending order, according to the prediction rate of the recommender system for them. \textbf{Initials of professional critics}: Wine Advocate --- Lisa Perotti Brown, Decanter --- Steven Spurrier, James Lawther, Beverley Blanning, and Jane Anson, Revue du Vin de France --- Olivier Poels, Hélène Durange, and Philippe Maurange,  Le Point --- Jacques Dupont, PersWijn --- Ronald DeGroot}}
\label{fig:correlationExperts}
\end{center}
\end{figure*}

\enlargethispage{2em}

Our analysis revealed that there are substantial differences in the correlation profiles of critics and amateurs, but also in the influence potential of different critics. To have better insight into how these differences are produced, we looked at the correlation profiles of different critics with other critics and amateurs. It can be easily observed that the  correlations of critics with other critics (purple beads) are much higher than correlations with amateur raters (orange beads). In fact, for most critics (possibly with the exception of Jacques Dupont writing for Le Point and Jane Anson) other critics are among the most correlated other individuals (the purple beads are on the far right of the correlation distribution).

\newpage

\subsection{Average recommender potential and influence of critics and amateurs}

\begin{figure*}[tbh!]
\begin{center}
\includegraphics[width=1\textwidth]{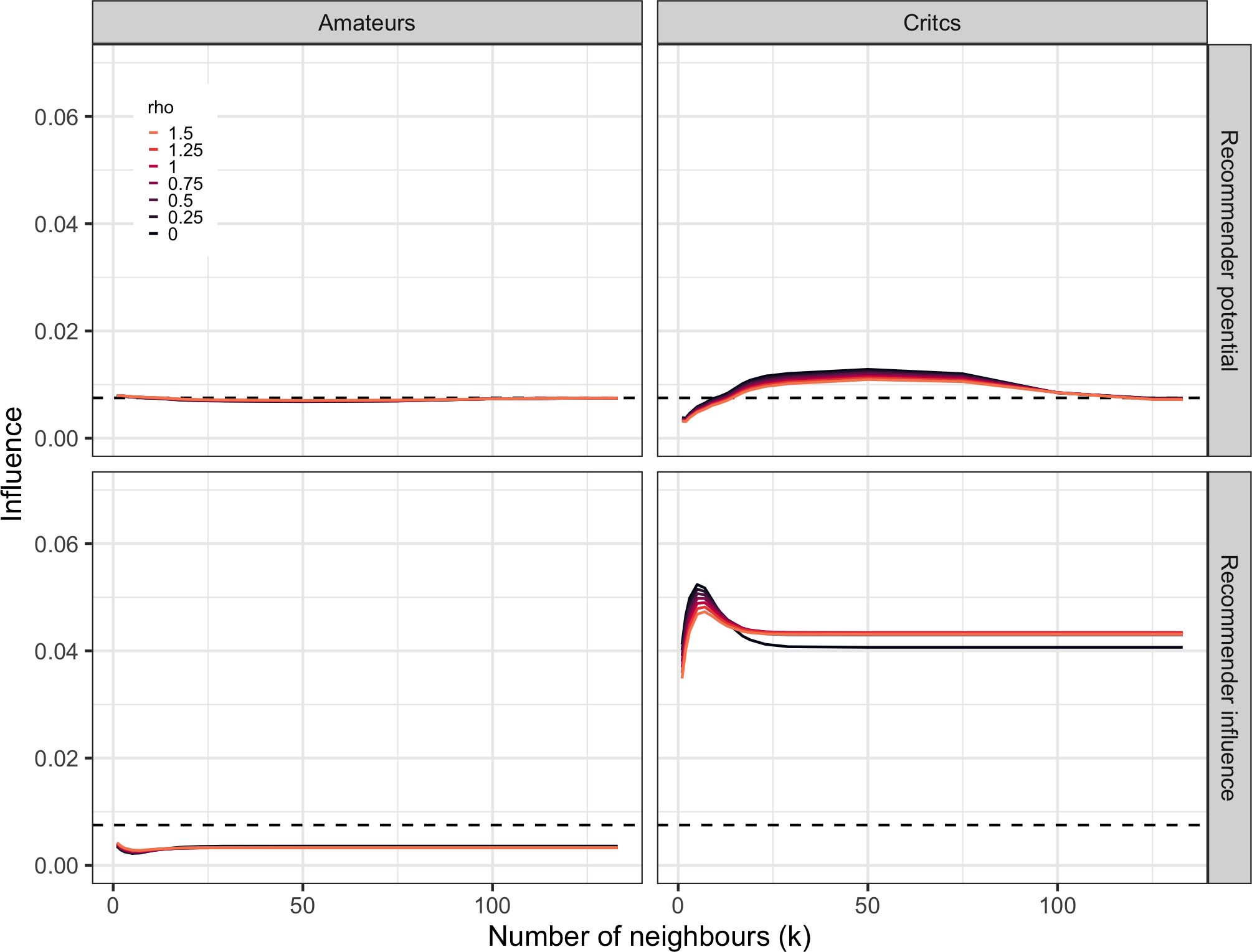}
\caption{\textbf{The average recommender potential and recommender influence of amateurs and critics as a function of the parameters $k$ and $\rho$. The horizontal dashed line represents the average recommender potential in the case where people consider the opinions of all other individuals using an equal weights strategy.}
} 
\label{fig:averageRecommenderPotential}
\end{center}
\end{figure*}

For small values of $k$ amateurs have a larger recommender potential than critics. This happens because the correlations among amateurs tend to be more dispersed and the people with the highest correlations with amateurs tend to be other amateurs. This picture changes as $k$ increases because the opinions of critics are consistently sought by the recommender system for larger $k$ values. In practice, the critics are substantially more influential than amateurs. This is because they are much more prolific raters, and they can more often supply their advice when the recommender system seeks it. Note that there are relatively small changes in the two groups as we vary the number of neighbors $k$ and the similarity sensitivity parameter $\rho$. 

\end{document}